\documentclass[final,3p]{elsarticle}
\pdfoutput=1
\usepackage{amsmath} 
\usepackage{amssymb}
\usepackage{textcomp}
\usepackage{graphicx}		
\usepackage{float}
\usepackage{color}
\usepackage{url}
\usepackage{array}
\usepackage{booktabs,ctable,multirow}
\usepackage{bm}
\usepackage[T1]{fontenc}

\DeclareFontFamily{U}{msb}{}
\DeclareFontShape{U}{msb}{m}{n}{ <5> <6> <7> <8> <9> gen * msbm
  <10> <10.95> <12> <14.4> <17.28> <20.74> <24.88> msbm10}{} 
\DeclareSymbolFont{AMSb}{U}{msb}{m}{n}
\DeclareMathSymbol{\I}{\mathalpha}{AMSb}{"49}
\DeclareMathSymbol{\N}{\mathalpha}{AMSb}{"4E}
\DeclareMathSymbol{\R}{\mathalpha}{AMSb}{"52}
\DeclareMathSymbol{\Z}{\mathalpha}{AMSb}{"5A}
\newcommand{\red}{\color[rgb]{1,0,0}}
\newcommand{\SF}{\bfseries \sffamily}

\newcommand{\bK}{\bm{K}}
\newcommand{\br}{\bm{r}}
\newcommand{\bx}{\bm{x}}

\newcommand{\bD}{\bm{D}}
\newcommand{\bI}{\bm{I}}
\newcommand{\bL}{\bm{L}}
\newcommand{\bP}{\bm{P}}
\newcommand{\bR}{\bm{R}}
\newcommand{\bW}{\bm{W}}
\newcommand{\bba}{\bm{\beta}}

\newcommand{\bsi}{\bm{\psi}}
\newcommand{\Bsi}{\bm{\Psi}}

\newcommand{\dt}{\Delta t}
\newcommand{\barre}{\rule[5mm]{\textwidth}{0.5pt}\\\vspace*{-10mm}~~}%
\usepackage[colorlinks=true]{hyperref}

\begin{document}
\begin{frontmatter}
\title{Exploring the Manifold of Seismic Waves:\\ Application to the Estimation of Arrival-Times}
\author[add1]{K.M. Taylor}
\author[add2]{M.J. Procopio}
\author[add2]{C.J. Young}  
\author[add3]{F.G. Meyer\corref{cor1}}
\ead{E-mail: fmeyer@colorado.edu}
\cortext[cor1]{Corresponding author}
\address[add1]{Department of Applied Mathematics,  University of Colorado at Boulder,
Boulder, CO}
\address[add2]{Sandia National Laboratories, Albuquerque, NM}
\address[add3]{Department of Electrical Engineering, University of Colorado at
Boulder, Boulder, CO}
\begin{abstract}
  We propose a new method to analyze seismic time series and estimate
  the arrival-times of seismic waves. Our approach combines two
  ingredients: the times series are first lifted into a
  high-dimensional space using time-delay embedding; the resulting
  phase space is then parametrized using a nonlinear method based on
  the eigenvectors of the graph Laplacian. We validate our approach
  using a dataset of seismic events that occurred in Idaho, Montana,
  Wyoming, and Utah, between 2005 and 2006. Our approach outperforms
  methods based on singular-spectrum analysis, wavelet analysis, and
  STA/LTA.
\end{abstract}
\end{frontmatter}
% ______________________________________________________________________________________________
\section{Introduction}
\label{introduction}
% ______________________________________________________________________________________________
\subsection{Estimation of arrival-times}
%______________________________________________________________________________________________
Seismic waves come in distinct bursts, or arrivals, corresponding to
different propagation paths through the earth. Arrival-times of
seismic waves are indispensable to the determination of the location
and type of the seismic event; the precise estimation of arrival-times
remains therefore a fundamental problem.  This paper addresses the
problem of estimating arrival-times of local seismic waves from a
seismogram. Several methods for estimating arrival-times use some
variants of the classic current-value-to-predicted-value ratio method
(e.g. \citep{Allen82,Panagiotakis08,Distefano06} and references
therein). The current value is a short term average (STA) of the
energy of the incoming data, while the predicted value is a long term
average (LTA), so the ratio is expressed as STA/LTA. This ratio is
constantly updated as new data flows in, and a detection is declared
when the ratio exceeds a threshold value.  When the signal and the
noise are Gaussian distributed, the STA/LTA method yields an optimal
detector that strikes the optimal balance between Type I and Type II
errors \citep{Freiberger63,Berger01}. As explained in
\cite{Persson03}, seismic waves are non-Gaussian, and therefore higher
order statistics (such as skewness and kurtosis) can be used to detect
the onset of seismic waves
\cite{Saragiotis02,Kuperkoch10,Galiana02}. The performance of a
detector can be improved by enhancing the signal transients relative
to the background noise. Several time-frequency and time-scale
decompositions have been proposed for this purpose (e.g.
\citep{Zhang03,Bardainne06,Withers98}, and references
therein). Advanced statistical methods can use training data (in the
form of seismograms labelled by an analyst). For instance, the
software developed at the Prototype International Data Center
(Arlington, VA) is based on a multi-layer neural network that uses
labelled waveforms in order to predict the types of waves of unseen
seismograms \citep{Wang02}.
%
%______________________________________________________________________________________________
\begin{figure}[H]
  \centerline{
    \includegraphics[width =30pc]{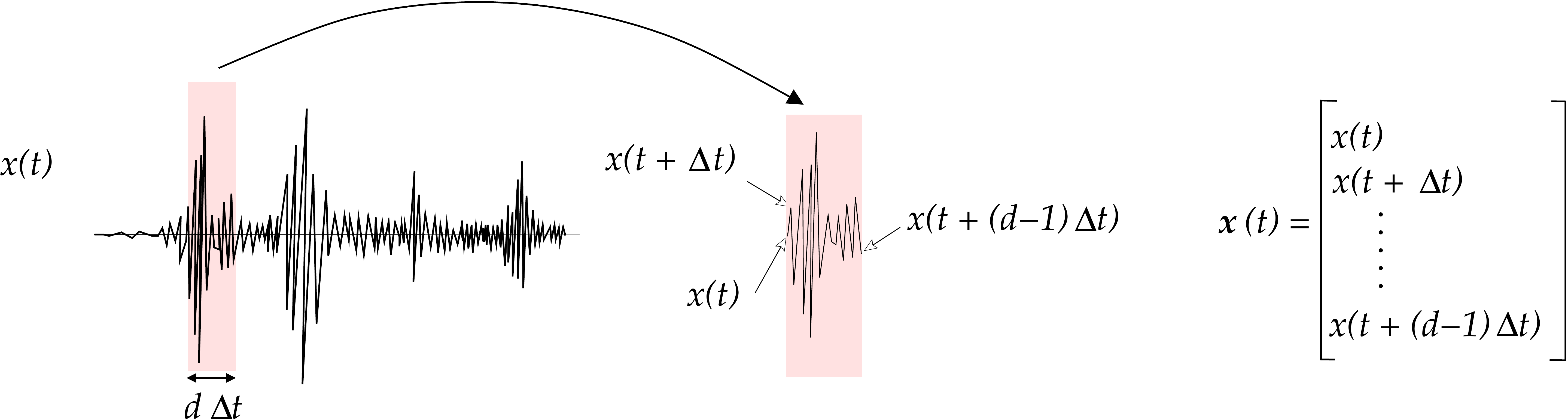}
}
    \caption{For each time $t$ we construct a patch $\bx(t)$ composed of
      $d$ equally-spaced samples of the seismogram.
    \label{patchdef}}
\end{figure}
%______________________________________________________________________________________________ 
\noindent 
%______________________________________________________________________________________________
\subsection{Local analysis of a time series: the concept of patch}
\begin{figure}[H]
  \centerline{
    \includegraphics[width = 27pc]{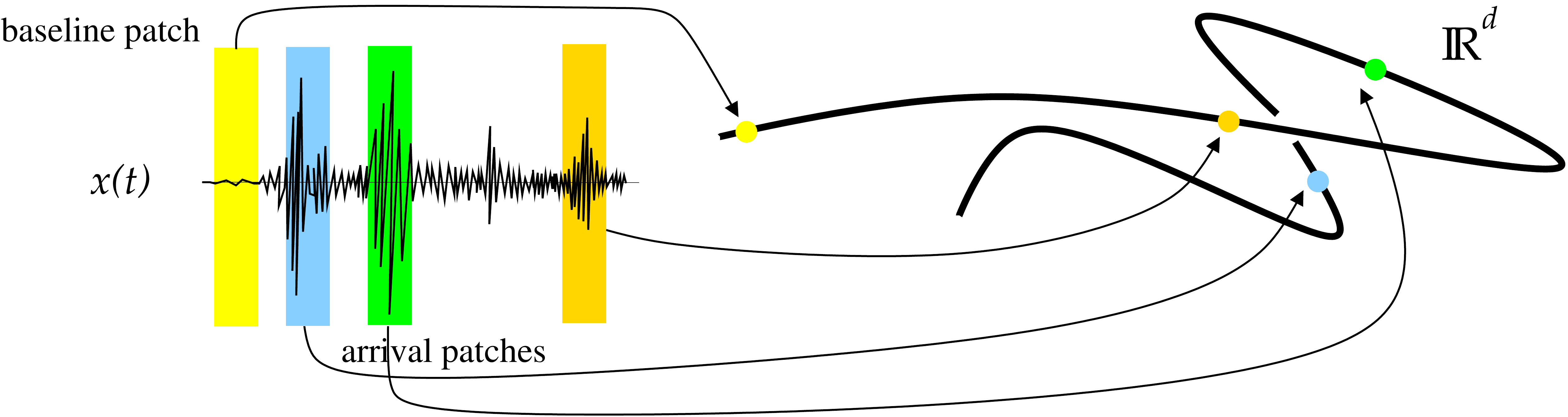}
}
    \caption{The trajectory $\left\{\bx(t), t \ge 0\right\}$ is a
      one-dimensional curve in $\R^d$. This 
      curve revisits already seen regions of $\R^d$ when the curent patch  $\bx(t)$
      is similar to a previously seen patch (e.g. the gold and blue patches).
      \label{patchspace}}
\end{figure}

%______________________________________________________________________________________________
In order to detect the arrival of a seismic wave, and estimate its
arrival-time, we propose to characterize the local dynamics of the
seismogram using a sliding window, or temporal {\em patch} (see
Fig.~\ref{patchdef}). Formally, let $x(t)$ be one of the components
(e.g. the $z$ component, or the first eigenmode after a singular value
decomposition of the three components) of a seismogram. Let $\dt$ be
the sampling period of the seismogram.  The {\em patch} $\bx(t)$ is
formed by collecting $d$ equally-spaced samples of $x(t)$ and stacking
them into a $d$-dimensional vector,
\begin{equation}
  \bx(t) = 
  \begin{bmatrix}
   x(t) & x (t + \dt) & \ldots & x (t + (d-1)\dt)
\end{bmatrix}^T
.  
\label{delayembed}
\end{equation}
While we generally think of $\bx(t)$ as a snippet of the original
signal over the interval $[t, t+d\dt)$ (see Fig.~\ref{patchdef}),
in this paper we will think about $\bx(t)$ as a vector in $d$
dimensions (see Fig. \ref{patchspace}). We call {\em patch-space} the
region of $\R^d$ formed by the patch trajectory $\left \{\bx(t), t \ge
  0\right\}$ (see Fig.~\ref{patchspace}). Our goal is to detect the
presence of a seismic wave in the patch $\bx(t)$ from the location of
the patch in patch-space. 

The concept of patches is equivalent to the concept of {\em time-delay
  coordinates} in the context of the analysis of a dynamical system
from the time series generated by an observable
\citep{Sauer91,Abarbanel93,Gilmore98}.  In this work, we offer a novel
perspective on the concept of time-delay coordinates by combining
several patch trajectories (from several seismograms) and computing a
nonlinear parametrization of the combined phase spaces defined by the
delay coordinates (\ref{delayembed}). This nonlinear parametrization
assigns to each patch $\bx(t)$ a small number of coordinates that
uniquely characterize the position of the patch in  patch
space, while providing the optimal separation between baseline patches
and arrival patches.
%______________________________________________________________________________________________
\subsection{Problem statement}
%______________________________________________________________________________________________
We are interested in detecting seismic waves and estimating the
arrival-time of each wave. We model the seismogram $x(t)$ as a sum of
two components
\begin{equation}
  x(t) = b(t) + w(t),
  \label{model}
\end{equation}
where $w(t)$ represents a seismic wave arriving at time $\tau$, and
$b(t)$ represents the baseline (or background) activity.  We assume
that $b(t)$ is a slowly varying signal with its energy uniformly
distributed over time. In contrast, we expect that $w(t)$ will be a
fast oscillatory transient localized around the arrival-time $\tau$
(see Fig. \ref{distance}). Our goal is to detect the seismic wave
$w(t)$, and estimate its onset $\tau$. The difficulty of the problem
stems from the fact that there is a large variability in the shape and
frequency content of the seismic waves $w(t)$.

We tackle this question by lifting the model (\ref{model}) into $\R^d$
using the time-delay embedding defined by~(\ref{delayembed}). As
explained in Sec. \ref{localdistances}, after time-delay embedding,
{\em baseline patches} that do not overlap with the seismic wave
$w(t)$ and only contain the baseline signal $b(t)$ become tightly
clustered along low-dimensional curves.  In contrast, {\em arrival
  patches} that include portions of the seismic wave $w(t)$ remain at
a large distance from one another, and are also at a large distance
from the baseline patches. The differential organization of the
baseline and arrival patches in $\R^d$ is the first ingredient of our
approach.

The second ingredient of our approach is provided by a parametrization
of patch-space that manages to cluster baseline patches and arrival
patches into two separate groups. This parametrization allows us%
%______________________________________________________________________________________________
%______________________________________________________________________________________________
\noindent  to
represent patches from $\R^d$, where $d$ is of the order of $10^3$
using only about $25$ coordinates. Finally, the last stage of our
approach consists in training a classifier to detect patches
containing seismic waves. The classifier uses the low-dimensional
parametrization of patch-space. 

In summary, the contribution of this paper is a novel method to
analyze seismograms and estimate arrival-times of seismic waves. Our
approach includes the following three steps:
\begin{enumerate}
\item Construction of patch-space by time-delay embedding of
  seismograms;
\item Low-dimensional parametrization of patch-space;
\item Construction of a classifier using the low-dimensional parametrization;
  detection of the presence of seismic waves and estimation of arrival-times.
\end{enumerate}
% ______________________________________________________________________________________________
\subsection{Outline}
% ______________________________________________________________________________________________
In the next section we explain how the properties of a waveform $x(t)$
will manifest themselves in terms of geometric properties of the patch
trajectory $\left\{\bx(t), t \ge 0\right\}$. In particular, we
estimate mutual distances between patches and we describe the
alignment of patches along low-dimensional subspaces. Our analysis is
performed assuming time is continuous (in Sec. \ref{discrete}, we
consider the discrete version of the problem). Furthermore, we assume
that the set of patches is constructed from a single seismogram. In
Sec. \ref{laplacian} we expand our discussion to the case where
patches are extracted from several seismograms collected at different
stations.  In Sec. \ref{lapeigs}, we construct a low-dimensional
parametrization of the discrete version of patch-space. We use this
low-dimensional parametrization in Sec. \ref{detect} to build a
classifier that learns to detect patches made up of seismic
waves. Finally, the performance of our approach is quantified in
Sec. \ref{results}.
%______________________________________________________________________________________________
\section{Patch-Space}
%______________________________________________________________________________________________
%______________________________________________________________________________________________
\subsection{Patch trajectories are  one-dimensional}
%______________________________________________________________________________________________
Given a seismogram $x(t)$, the patch $\bx(t)$ extracted at time $t$ is
a vector in $\R^d$. As $t$ evolves, the set of points $\left\{\bx(t),
  t\ge 0\right\}$ forms the patch trajectory. If $x(t)$ is a smooth
function of time (has $s$ derivatives), then the patch trajectory is a
smooth one-dimensional curve in $\R^d$, and the dimension of the curve
(one) is independent of the patch size $d$. Indeed, for
$k=0,\ldots,d-1$ the maps
\begin{equation}
t \longrightarrow x(t + k \dt)
\end{equation}
are all smooth. Therefore each of the coordinates of $\bx(t)$ is a
smooth function of $t$, and the map 
\begin{equation}
t \longrightarrow \bx(t)
\end{equation}
is a smooth map from $\R$ to $\R^d$ and thus is a
one-dimensional curve in $\R^d$.
%______________________________________________________________________________________________
\subsection{Mutual Distance Between Two Patches
\label{localdistances}}
\begin{figure}[H]
\centerline{
    \includegraphics[width = 16pc]{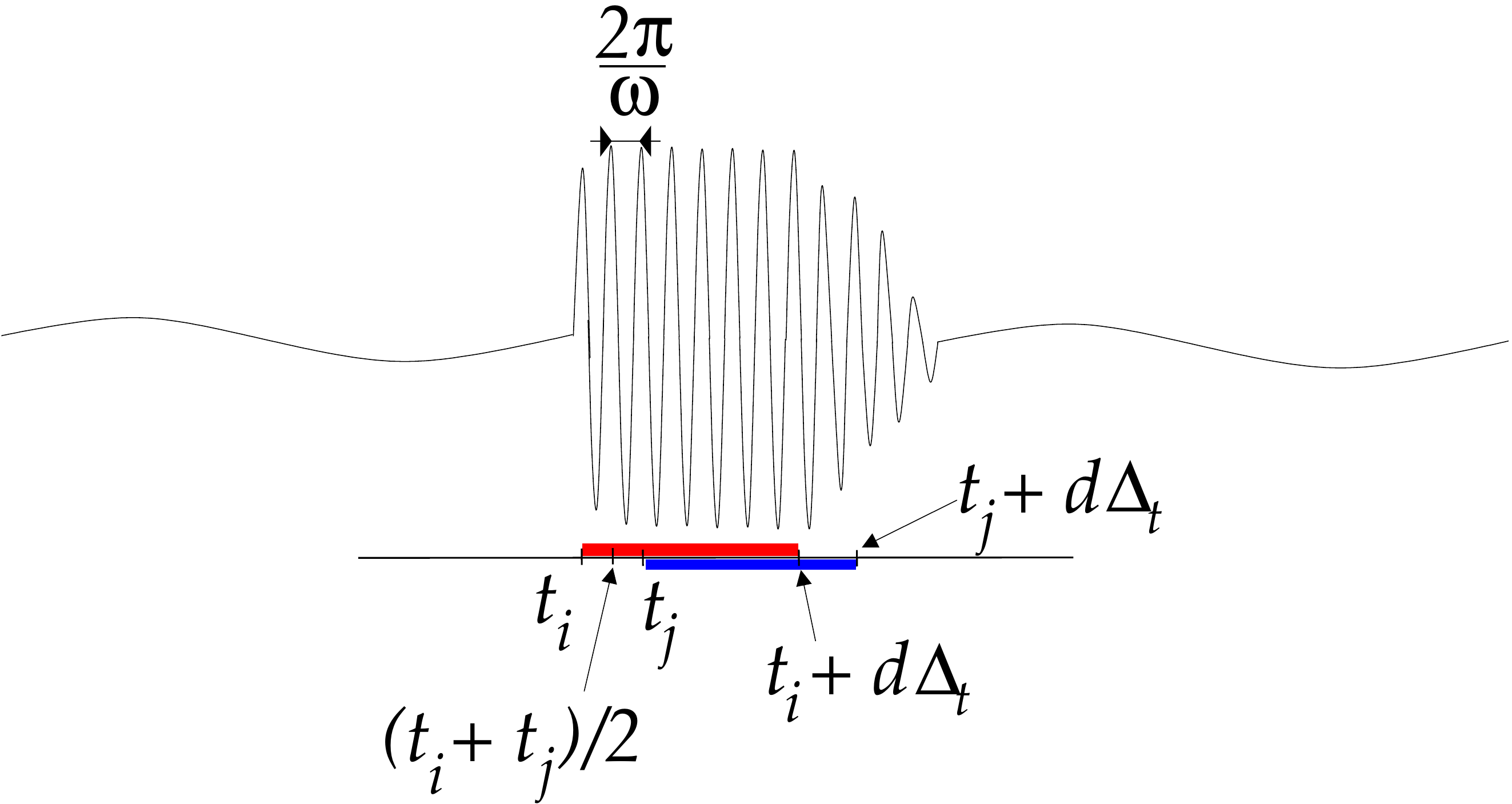}
}
\caption{The two patches (in red and in blue) contain part of the seismic wave $w(t)$.
  \label{distance}}
\end{figure}

% ______________________________________________________________________________________________
In the following, we assume that the seismogram is described by the model (\ref{model})
and we study the Euclidean distance between any two patches $\bx(t_i)$
and $\bx(t_j)$ extracted at times $t_i$ and $t_j$.
We first consider the case where both patches come from the baseline
part of the signal. In this case, we assume $w(t) = 0$ over the
intervals $[t_i, t_i + d\dt)$ and $[t_j, t_j +d\dt)$, and we have
\begin{equation*}
  \|\bx(t_i)-\bx(t_j)\|^2 =  \sum_{k = 0}^{d-1}\left(b(t_i+k\dt)-b(t_j+k\dt)\right)^2.
\end{equation*}
If the baseline signal varies slowly, then we have
$\left |b(t_i+k\dt)-b(t_j+k\dt)\right| \approx 0, \; (k=0,\ldots,d-1)$,
and therefore
\begin{equation*}
  \|\bx(t_i)-\bx(t_j)\|^2  = \sum_{k = 0}^{d-1} 
  \left (b(t_i+k\dt)-b(t_j+k\dt)\right)^2 \approx 0. 
\end{equation*}
We now consider the case where one patch, $\bx(t_i)$ (without loss of
generality), is part of a seismic wave $w(t)$, whereas the other
patch, $\bx(t_j)$, comes from the baseline part. We have
\begin{eqnarray*}
  \|\bx(t_i)-\bx(t_j)\|^2 &=& \sum_{k = 0}^{d-1}\left(w(t_i+k\dt)
    + b(t_i+k\dt)-b(t_j+k\dt)\right)^2\\
  &=& \sum_{k = 0}^{d-1} 
  w^2(t_i+k\dt) + 
2\sum_{k = 0}^{d-1}   w(t_i+k\dt)  \left
  (b(t_i+k\dt)-b(t_j+k\dt)\right)\\
&&+ 
  \sum_{k = 0}^{d-1}  \left (b(t_i+k\dt)-b(t_j+k\dt)\right) ^2.
\end{eqnarray*}
Again, we can assume that for each $k$, $\left |b(t_i+k\dt)-b(t_j+k\dt)\right|
\approx 0$, and thus 
\begin{equation*}
  \sum_{k = 0}^{d-1} 
  w(t_i+k\dt)
  \left (b(t_i+k\dt)-b(t_j+k\dt)\right)
  \approx 0.
\end{equation*}
As before, we have $\sum_{k = 0}^{d-1} \left (b(t_i+k\dt)-b(t_j+k\dt)\right)^2   \approx 0$ . 
We conclude that 
\begin{equation}
  \|\bx(t_i)-\bx(t_j)\|^2  = \sum_{k = 0}^{d-1} 
  w^2(t_i+k\dt).
\label{energy1} 
\end{equation}
This sum measures the energy of the (sampled) seismic
wave over the interval $[t_i, t_i + d \dt)$.  Because the patch size,
$d\dt$ , is chosen so that $w (t)$ oscillates several times over the
patch (see Fig. \ref{distance}), the interval $[t_i, t_i + d \dt)$ is
comprised of several wavelengths of $w(t)$, and the energy
(\ref{energy1}) is usually large.

Finally, we consider the case where both patches contain part of the
seismic wave $w(t)$ (see Fig. \ref{distance}),
\begin{eqnarray*}
  \|\bx(t_i)-\bx(t_j)\|^2 
 & = &  \sum_{k = 0}^{d-1}  \left (b(t_i+k\dt)-b(t_j+k\dt)\right) ^2   
  +   \sum_{k = 0}^{d-1}   \left(w(t_i+k\dt) - w(t_j+k\dt) \right)^2  \\
  & + & 2\sum_{k = 0}^{d-1} 
  \left (w(t_i+k\dt)-w(t_j+k\dt) \right)
  \left (b(t_i+k\dt)-b(t_j+k\dt)\right).
\end{eqnarray*}
If we assume that the baseline signal varies slowly over time, then we
have again
\begin{equation}
  \|\bx(t_i)-\bx(t_j)\|^2 
  \approx 
  \sum_{k = 0}^{d-1} 
  \left (w(t_i+k\dt) - w(t_j+k\dt) \right)^2 .
\label{diffwave}
\end{equation}
%
%______________________________________________________________________________________________
%______________________________________________________________________________________________
\noindent The sum (\ref{diffwave}) measures the energy of the
difference between two overlapping sections of the seismic wave $w(t)$,
sampled every $\dt$ (see Fig. \ref{distance}). In order to estimate
the size of this energy, we approximate the seismic wave with a cosine
function (see Fig. \ref{distance}), $w(t) = \cos (\omega t)$, where
the frequency $\omega$ corresponds to the peak of the short-time
Fourier transform of $w(t)$ around $\tau$. In this case, we have
\begin{eqnarray*}
  w(t_i+k\dt) - w(t_j+k\dt) & =& 
  \cos \left\{\omega (t_i + k \dt)\right\} -
  \cos \left\{\omega (t_j + k \dt)\right\} \\
  & =& 2 \sin \left \{\omega (t_j - t_i)/2 \right\} 
  \sin \left\{\omega \left[(t_i + t_j)/2 + k \dt\right]\right\},
\end{eqnarray*}
and the sum (\ref{diffwave}) becomes
\begin{equation}
  \sum_{k = 0}^{d-1} 
  \left \{w(t_i+k\dt) - w(t_j+k\dt) \right\}^2 
  =  
  4 \sin^2 \left \{\omega (t_j - t_i)/2 \right \} 
  \sum_{k = 0}^{d-1} 
  \sin^2 \left\{\omega \left[(t_i + t_j)/2 + k \dt\right]\right\}.
\label{diff-wave2}
\end{equation}
The sum in the right-hand side of (\ref{diff-wave2}) can be written as
\begin{equation}
\sum_{k = 0}^{d-1} 
\sin^2 \left\{\omega \left[(t_i + t_j)/2 + k\dt\right]\right\}
= \sum_{k = 0}^{d-1} 
\cos^2 \left\{\omega \left[(t_i + t_j)/2 + \frac{\pi}{2\omega} + k\dt\right]\right\}
\label{energy}
\end{equation}
The right-hand side of (\ref{energy}) is the energy of the cosine
function sampled every $\dt$ on an interval starting at $(t_i+t_j)/2 +
\pi/2\omega$ of length $d\dt$. As explained above $d \dt \gg 2\pi/\omega$, 
and thus the energy (\ref{energy}) is measured over several wavelengths and is
therefore large. Finally, given a patch
starting at time $t_i$, all patches starting at time $t_j$, where $t_j
= t_i + (q +1/2) 2\pi/\omega$, for some $q =0,1,\ldots$ will satisfy
$\sin^2 \left (\omega (t_j - t_i)/2 \right) = 1$. We conclude that
given a patch starting at time $t_i$, there are many choices of $t_j$
such that the sum in (\ref{diff-wave2}), and therefore the norm in
(\ref{diffwave}), are large.

In summary, we expect the mutual distance between patches extracted
from the baseline signal to be small, while the mutual distance
between baseline and arrival patches will be large.  Moreover, we also
expect that two arrival patches will often be at a large distance of
one another.
% ______________________________________________________________________________________________
\subsection{Global Alignment of the Patches}
%______________________________________________________________________________________________
We have seen that, given a seismogram $x(t)$, the patch trajectory
$\left \{\bx(t), t \ge 0\right\}$ is a one-dimensional curve in
$\R^d$. We could imagine that this curve would be wildly scattered all
over $\R^d$, exploring every part of the space. In fact, this is not
the case: if the seismogram has bounded derivatives up to order $s$,
then the patch trajectory will lie near the
intersection of $s$ hyperplanes. In other words, the one-dimensional
curve does not explore the entire space, but stays inside a subspace
of dimension $d-s$

We first observe that we can compute numerical estimates of the first
$s \le d-1$ derivatives from the $d$ coordinates of the patch $\bx(t)$. We
can therefore quantify the local regularity of the function $x(t)$
over the time window $[t,t+d\dt)$ from the knowledge of $\bx(t)$.
This observation was at the origin of the first embedding theorems
\citep{Packard80} that did not use the notion of time-delay coordinates --
which is equivalent to our%
%
%______________________________________________________________________________________________
\begin{figure}[H]
\centerline{
    \includegraphics[width = 10pc]{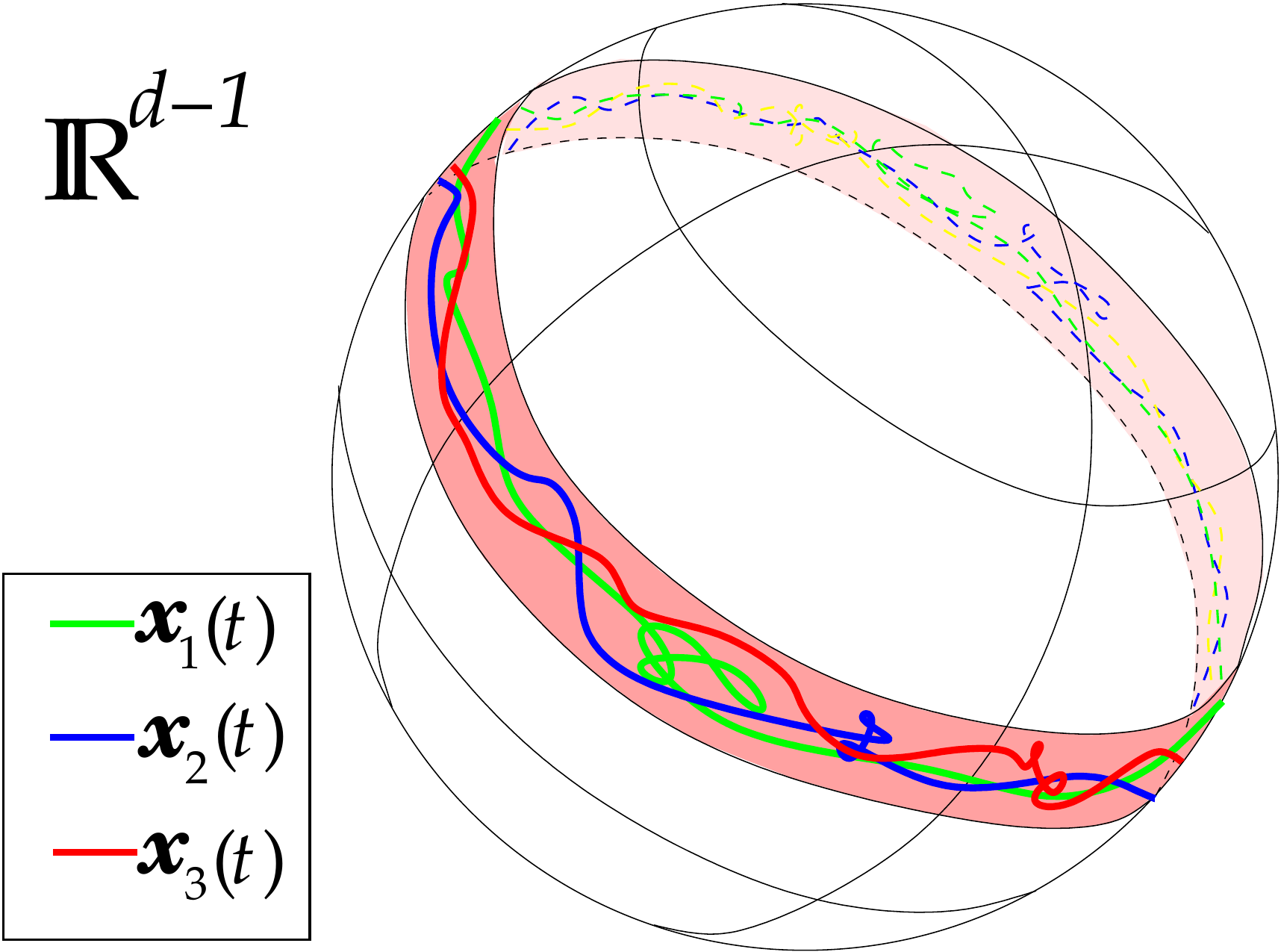}
}
\caption{The patch trajectories $\bx_1(t),\bx_2(t)$ and $\bx_3(t)$ 
  belong to  a common low-dimensional submanifold of the
  $d-2$-dimensional unit sphere in $\R^{d-1}$.
  \label{manypatches}}
\end{figure}
%______________________________________________________________________________________________
\noindent notion of patch -- but involved the
differential coordinates $d^px/dt^p, p =0,\ldots, d-1$.  We consider a
forward finite difference scheme that approximates the derivative of
order $p$ ($1 \le p \le s-1$) at time $t$,
\begin{equation}
\frac{d^px}{dt^p} \approx  \frac{1}{(\dt)^p}\sum_{j=0}^{p} a_j x(t+ j \dt).
\label{difference}
\end{equation}
Obviously, more accurate schemes can be obtained using central
differences; we use a forward scheme here to simplify the
discussion. The finite difference (\ref{difference}) is a linear
combination of the first $p+1$ coordinates of the patch $\bx(t)$. We
assume that $d^px/dt^p$ is bounded by a small constant over an
interval $U$. Consequently, the finite difference (\ref{difference})
is small, and we have
\begin{equation*}
\left |\sum_{j=0}^{p} a_j x(t+ j \dt) \right | \approx 0,\qquad t\in U.
\end{equation*}
This last statement can be translated as follows: the distance of the
patches $\left\{\bx(t), t \in U\right\}$ to the hyperplane of $\R^d$
defined by the vector
\begin{equation}
  \begin{bmatrix}
    a_0 & a_1 & \cdots & a_p & 0 & \cdots & 0
  \end{bmatrix}^T \in \R^d.
  \label{hyperplane}
\end{equation}
is small. We conclude that if the function $x(t)$ has $s$ bounded
derivatives for $t \in U$, then the trajectory of $\bx(t)$, for $t \in
U$, lies near the intersection of $s$ hyperplanes, each of which is
defined by a set of coefficients of the form (\ref{hyperplane}).
% ______________________________________________________________________________________________
\subsection{Normalization of Patch-Space}
\label{normalization}
% ______________________________________________________________________________________________
We now consider the following question: if we want to use seismograms
from different stations to learn the general shape of a seismic wave
(see e.g. \cite{Aster00}), how should we normalize the seismograms?
The magnitude of an earthquake, which characterizes its damaging
effect, is defined as a logarithmic function of the radiated energy
\citep{BenZion08}. The radiated energy can be estimated by integrating
the velocity associated with the displacement measured by seismograms
\citep{BenZion08}. A logarithmic normalization \citep{BenZion08} would
make it possible to account for the large variability in the energy
and would allow us to compare seismograms from different stations or
from different events. We favor an equivalent normalization that
consists in rescaling each patch by it energy.  For every patch
$\bx(t)$, we first remove any slowly varying drift by removing the
mean of $x(t)$  over the interval $[t, t+ d\dt)$, $  \overline{x}(t) =
\left(\sum_{k=0}^{d-1}  x (t + k\dt) \right)/d$, and we
compute the centered patch
\begin{equation*}
  \bx_0(t) = 
  \begin{bmatrix}
    x(t ) - \overline{x} (t) &  \ldots& 
    x(t + (d-1) \dt) - \overline{x} (t)
  \end{bmatrix}^T.
\end{equation*}
Finally, we project the centered patch $\bx_0(t)$  on  the
unit sphere and define the normalized patch
\begin{equation}
 \frac{\bx_0(t)}{\|\bx_0(t) \|}.
\label{onthesphere}
\end{equation}
The normalized patch (\ref{onthesphere}) characterizes the local
oscillation of $x(t)$ in a manner that is independent of changes in
amplitude and of any slow drift of the seismogram. Geometrically, the
trajectory of the normalized patch is a curve on the $d-2$ dimensional
unit sphere in $\R^{d-1}$ (see Fig.~\ref{manypatches}). Indeed, after
subtracting the mean $\overline{\bx}(t)$, the patch lies on the
hyperplane of $\R^d$ defined by $\sum_{i=1}^d x_i = 0$. After the
normalization (\ref{onthesphere}), the normalized patch lives at the
intersection of the unit sphere in $\R^d$, and the hyperplane
$\sum_{i=1}^d x_i = 0$. This intersection is also a unit sphere, but in
$\R^{d-1}$. In the remaining of the paper we assume that each patch
$\bx(t)$ has been normalized according to (\ref{onthesphere}).
% ______________________________________________________________________________________________
\subsection{The Embedding Dimension}
% ______________________________________________________________________________________________
The patch size is determined by the number, $d$, of time-delay
coordinates. The selection of $d$ can be guided by several algorithms
that have been proposed in the context of the analysis of nonlinear
dynamical systems from time series \citep{Judd98,Kennel02,Small04}. In
practice, if $d$ is chosen too large, then our understanding of the
geometry of patch-space becomes restricted by the number of
patches available. Indeed, the number of patches is limited by the
number of time samples; but we need a number of patches that grows
exponentially with the dimension, $d$ ,of patch-space to properly
estimate the distribution of the patches \citep{Scott92}.
%______________________________________________________________________________________________
\section{A new parametrization of patch-space}
\label{laplacian}
%______________________________________________________________________________________________
\subsection{From a single seismogram  to several seismograms}
%______________________________________________________________________________________________
Because we learn to detect arrivals using more than a single seismic
trace (as in \cite{Aster00}, for instance) we need to understand the
structure of patch space when patches come from different seismograms
acquired at different stations. After normalization, all the patches
live on the $d-2$-dimensional unit sphere in $\R^{d-1}$, and are
therefore characterized by $d-2$ coordinates. Each seismogram gives
rise to a one-dimensional trajectory on the unit sphere (see
Fig.~\ref{manypatches}). We expect that the patch trajectories created
by different seismograms will remain close to one another, and will
not be spread across the unit sphere. Our experiments confirm that the
trajectories belong to a $m$-dimensional submanifold embedded in the
$d-2$-dimensional unit sphere, where $m \ll d-2$.
%______________________________________________________________________________________________
\subsection{From continuous to discrete patch-space
\label{discrete}}
%______________________________________________________________________________________________
In practice, each seismogram is sampled with the sampling period
$\dt$, and we obtain a time series
\begin{equation*}
  x_i = x(t_i), \qquad \text{where} \quad t_i = i \dt.
\end{equation*}
The uniform sampling (in time) of the seismogram leads to a non
uniform sampling (in space) of the one-dimensional trajectory
$\bx(t)$.  We define the discrete patch by
\begin{equation}
  \bx_i =
\begin{bmatrix}
x_i & x_{i+1} & \ldots & x_{i+(d-1)}
\end{bmatrix}^T.
\end{equation}
In order to identify patches that contain seismic waves, we want to
characterize the $m$-dimensional submanifold of patch
trajectories. Formally, we seek a smooth parametrization of the set of
patch trajectories that uses the minimum number of parameters --
theoretically only $m$. An answer to this question is provided by the
computation of the principal components of the set of patches
$\left\{\bx_i, i=0,\ldots\right\}$. This method, known as
singular-spectrum analysis \citep{Broomhead86,Vautard89}, has been
used to characterize geophysical time series
\citep{Sharma93,Gamiz02}. Unfortunately, unless the patches lie along
a low-dimensional linear subspace, many (typically more than $m$)
principal components will be required to capture the curvature and
torsion of the patch trajectories. This problem is quite severe since
the part of patch-space that corresponds to arrivals exhibits high
curvature. Our quest for a parametrization of the submanifold of patch
trajectories is similar to the problem of reconstructing a phase space
based on time-delay coordinates.  Indeed, we can consider that
seismograms are observables from the nonlinear dynamical system that
is at the origin of the seismic waves. The authors in
\citep{Delauro08} and \citep{Demartino04} show that patches extracted
from volcanic tremors evolve around a low-dimensional attracting
manifold (see also \citep{Chouet91,Godano96,Yuan04,Konstantinou02b})
The embedding dimension of the phase space reconstructed from
Strombolian tremors was estimated in \citep{Delauro08} to be around
five (see also \citep{Demartino04} and \citep{Konstantinou02}). Our
approach provides%
%______________________________________________________________________________________________
\begin{figure}[H]
  \centerline{
    \includegraphics[width =20pc]{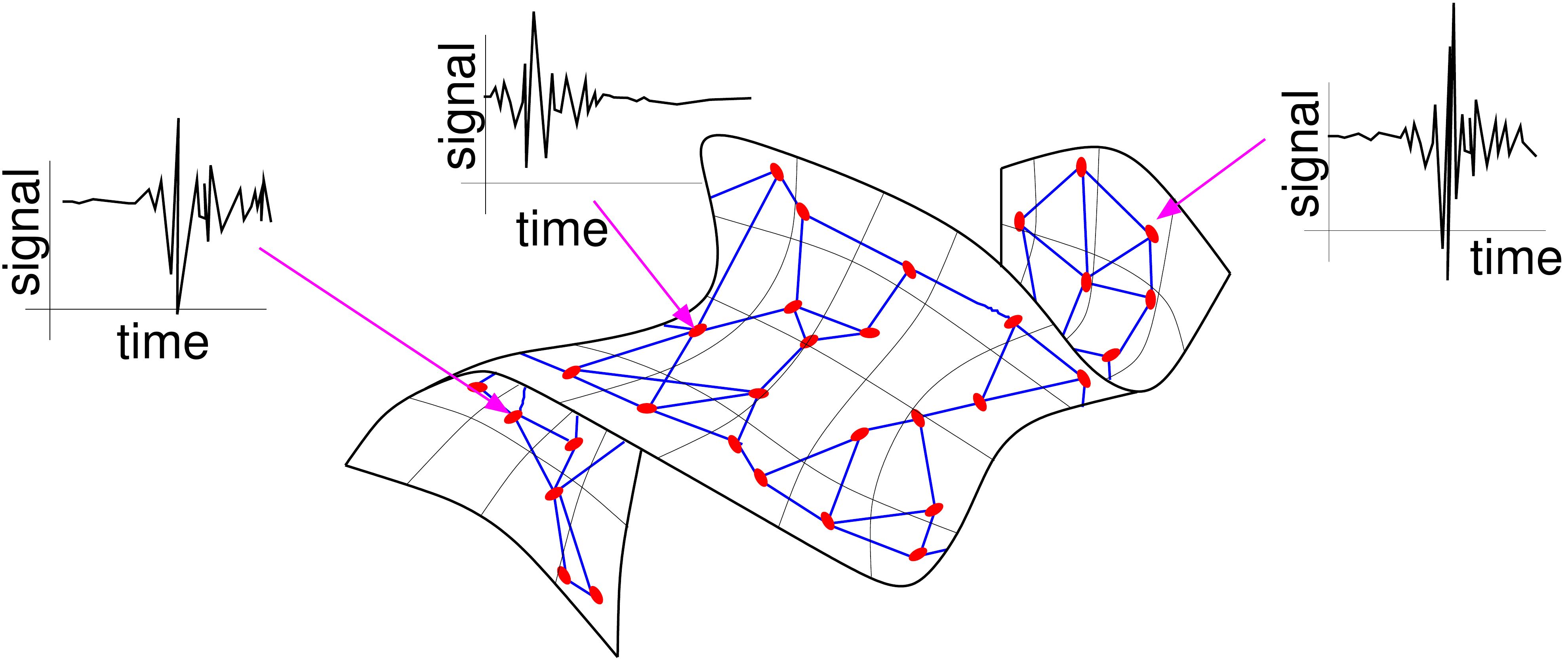}
}
    \caption{The patch graph: each node is a patch; nodes are
      connected if patches are similar.
   \label{patchgraph}}
\end{figure}
%______________________________________________________________________________________________ 
\noindent a new perspective on this question by directly
constructing a smooth parametrization of the set of phase spaces
associated with the different seismograms. Our hypothesis, confirmed
by our experiments, is that the vectors $\bx_i$ live close to a
low-dimensional submanifold of the unit sphere in $\R^{d-1}$.
% ______________________________________________________________________________________________
\subsection{Nonlinear Parametrization of Patch-Space via the
  Eigenvectors of the Laplacian}
\label{lapeigs}
%______________________________________________________________________________________________
%______________________________________________________________________________________________
\subsubsection{The patch graph: a network of patches }
%______________________________________________________________________________
Our plan is to assemble a global parametrization $\Bsi$  of the submanifold of
patches from the knowledge of the pairwise distances between patches.
In principle, we should measure the geodesic distances between patches
along the underlying submanifold. Unfortunately, the only distances
available to us are Euclidean distances. This issue can be resolved by
observing that the Euclidean and the geodesic distances between two
patches are very similar if the patches are in close proximity. We
therefore use only distances between neighboring patches, and
disregard distances between faraway patches.

We describe patch-space with a graph $G$ that is constructed as
follows.  We assume that we have access to $N$ patches $\bx_i, i
=1,\ldots,N$ extracted from several seismograms recorded at different
stations. All the patches have been properly normalized, as explained
in Sec. \ref{normalization}. Each patch $\bx_i$ becomes the vertex
$\bx_i$ of the graph\footnote{We slightly abuse notation here: $\bx_i$
  is patch as well as a vertex on the graph.}. Edges between vertices
quantify the proximity between patches. Each vertex $\bx_i$ is
connected to its $\nu_{N\negmedspace N}$ nearest neighbors according
to the Euclidean distance $\| \bx_i - \bx_j\|$.  When the vertices
$\bx_i$ and $\bx_j$ are connected by an edge $\{i,j\}$ we write $\bx_i
\sim \bx_j$. The weight $W_{i,j}$ on the edge $\{i,j\}$ measures the
similarity between the patches $\bx_i$ and $\bx_j$ and is defined by
\begin{equation}
W_{i,j} = 
\begin{cases}
e^{\displaystyle - \|\bx_i-\bx_j\|^2/ \sigma^2}, 
& \text{if $\bx_i$ is connected to $\bx_j$},\\
0 & \text{otherwise.}
\end{cases}
\label{weight}
\end{equation}
The scaling factor $\sigma$ modulates our definition of proximity. If
$\sigma \gg \max_{i,j} \|\bx_i - \bx_j\|$, then for all edges
$\{i,j\}$ we have $W_{i,j} \approx 1$.  This choice of $\sigma$ blurs
the distinction between baseline and arrival patches by pretending
that all patches are similar to one another $(W_{i,j} \approx 1)$. On
the other hand, if $\sigma \approx 0$, then $W_{i,j} \approx 0$ for
all edges $\{i,j\}$ such that $\|\bx_i - \bx_j \|> 0$. Only if $\|
\bx_i - \bx_j\| \approx 0$ do we have $W_{i,j} \neq 0$. This choice of
$\sigma$ accentuates the differences between patches by pretending
that all patches are different the minute they are slightly
different. Obviously, this choice of $\sigma$ is very sensitive to any
noise existing in the data.  The weighted graph ${G}$ is fully
characterized by the $N \times N$ weight matrix $\bW$ with entries
$W_{i,j}$. We also define the diagonal degree matrix $\bD$ with
entries $D_{i,i} = \sum_{j}W_{i,j}$.
%______________________________________________________________________________________________
\subsubsection{We trust only local distances between patches}
%______________________________________________________________________________________________
The construction of the parametrization of patch-space is guided by
the following two principles:
\begin{itemize}
\item Distances between patches connected by an edge are small and
  one can approximate their geodesic distance by their Euclidean
  distance. This allows one to measure local geodesic distance without an
  existing knowledge of the underlying submanifold of patch
  trajectories.
\item The new parametrization of patch-space assembles the different
  constraints provided by the local geodesic distances into a set of
  global coordinates that vary smoothly over the submanifold.
\end{itemize}
Only an isometry will preserve exactly Euclidean distances between
patches, and the isometry which is optimal for dimension reduction is
given by PCA. However, as shown in the experiments, the coordinates
provided by PCA are unable to capture the nonlinear structures formed
by patch-space. We propose therefore to seek a sequence of functions
$\bsi_1, \bsi_2, \ldots$ that will become the new coordinates of each
patch.
%______________________________________________________________________________________________
\subsubsection{The new coordinate functions: the eigenfunctions of the Laplacian}
%______________________________________________________________________________________________
We define  each coordinate function $\bsi_k$ as the solution to the following
minimization problem
\begin{equation}
\min_{\bsi_k}
\frac{\sum_{{\bx_i\sim \bx_j}} W_{i,j}\left(\bsi_k(\bx_i) - \bsi_k (\bx_j) \right)^2}{\sum_i D_{i,i}\bsi_k^2(\bx_i)},
\label{rayleigh}
\end{equation}
where $\bsi_k$ is orthogonal to the previous functions $\{\bsi_0,
\bsi_1,\cdots,\bsi_{k-1}\}$,
\begin{equation}
\langle \bsi_k, \bsi_j \rangle = \sum_{i=1}^N
D_{i,i}\bsi_k(\bx_i)\bsi_j(\bx_i) = 0\quad\quad (j=0,\ldots,k-1).
\label{ortho}
\end{equation}
The numerator of the Rayleigh ratio (\ref{rayleigh}) is a weighted sum
of the gradient of $\bsi_k$ measured along the edges $\{i,j\}$ of the
graph; it quantifies the average local distortion introduced by the
map $\bsi_k$. The distortion is measured locally: if $\bx_i$ and
$\bx_j$ are far apart, then $W_{i,j} \approx 0$, and the difference
$(\bsi_k(\bx_i) - \bsi_k (\bx_j))$ does not contribute to the sum
(\ref{rayleigh}).  The denominator provides a natural
normalization. The constraint of orthogonality (\ref{ortho}) to the
previous coordinate functions guarantees that the coordinates
$\bsi_0,\bsi_1,\ldots$ describe the dataset with several resolutions:
if $\langle \bsi_k, \bsi_j \rangle = 0$ then $\bsi_k$ experiences more
oscillations on the dataset than the previous $\bsi_j$. Intuitively,
$\bsi_k$ plays the role of an additional digit that describes the
location of $\bx_i$ with more precision.  It turns out \citep{Chung97}
that the solution of (\ref{rayleigh},\ref{ortho}) is the solution to
the generalized eigenvalue problem,
\begin{equation}
\left(\bD - \bW\right) \bsi_k = \lambda_k \bD \bsi_k, \qquad \qquad k =0,1,\ldots
\label{geneigenprob}
\end{equation}
The first eigenvector $\bsi_0$, associated with $\lambda_0 =0$, is
constant, $\bsi_0 (\bx_i) =1, i=1,\ldots,N$; it is therefore not used.
Finally, the new parametrization  $\Bsi$ is defined by
\begin{equation}
 \bx_i  \mapsto  \Bsi(\bx_i) = 
\begin{bmatrix}
\bsi_1(\bx_i) & \bsi_2(\bx_i) & \ldots & \bsi_m(\bx_i) 
\end{bmatrix}^T.
\label{embed}
\end{equation}
The matrix $\bP = \bD^{-1}\bW$ is a row-stochastic matrix, associated
with a Markov chain on the graph, and the matrix $\bL = \bI - \bP$
that appears in (\ref{geneigenprob}) is known as the graph Laplacian.
The idea of parametrizing a manifold using the eigenfunctions of the
Laplacian can be traced back to ideas in spectral geometry
\citep{Berard94}, and has been developed extensively by several groups
recently \citep{Belkin03,Coifman06b,Coifman08,Jones08}. The
construction of the parametrization is summarized in Fig.~\ref{algo1}
(see also \cite{Saito08} for a version that can be implemented with
fast algorithms). Unlike PCA, which yields a set of vectors on which
to project each $\bx_i$, this nonlinear parametrization constructs the
new coordinates of $\bx_i$ by concatenating the values of the
$\bsi_k$, $k=1,\cdots,m$ evaluated at $\bx_i$, as defined in
(\ref{embed}).
%______________________________________________________________________________________________
\subsubsection{How many new coordinates do we need?}
% ______________________________________________________________________________________________
Our goal is to construct the most parsimonious parametrization of
patch-space with the smallest number $m$ of coordinates. We
expect that if $m$ is too small, then the new parametrization will not
describe the data with enough precision, and the detection of seismic
waves will be inaccurate. On the other hand, if $m$ is too large, then
some coordinates will be mainly describing the noise in the dataset
(and not adding additional information), and the classification
algorithm will overfit the training data. The experiments confirmed
that $m=25$ yields the optimal detection of seismic waves -- and
performed better than $m=50$ -- even when as many as $d=1024$
time samples are included in each patch. Clearly, this approach
results in a very significant reduction of dimensionality.
%_________________________________________________________________
\begin{figure}[H]
{\flushleft \SF Algorithm 1: Parametrization of patch-space}\\
  \barre
{\raggedright \SF Input}: \\
\begin{itemize}
\item  [] set of $N$ patches, $\bx_i, i=1,\cdots,N$; $m$: dimension of the embedding 
\item []   $\sigma$: width of the kernel for the graph; $\nu_{N\negmedspace  N}$: number of neighbors of each patch;
\end{itemize}
{\flushleft \SF Algorithm:}
    \begin{enumerate}
    \item construct the graph defined by the $\nu_{N\negmedspace N}$ nearest neighbors of each $\bx_i$
    \item compute the weight matrix $\mathbf W$, and the degree matrix $\mathbf D$
    \item compute the $m$ eigenvectors $\bsi_1,\ldots,\bsi_m$ of ${\mathbf
        D}^{-\frac{1}{2}}{\mathbf W}{\mathbf D}^{-\frac{1}{2}}$ 
    \end{enumerate}
{\flushleft \SF Output:}  $m$ coordinates 
    $\Bsi(\bx_i) = 
    \begin{bmatrix}
      \bsi_1(\bx_i) & \bsi_2(\bx_i) & \ldots & \bsi_m(\bx_i) 
    \end{bmatrix}^T.$

  \barre
  \caption{Construction of the embedding
    \label{algo1}}
\end{figure}
%_________________________________________________________________ 
\subsection{The case of a single station and a single event: a
  dynamical system connection}
%______________________________________________________________________________________________
We consider in this section the simpler situation where patch space is
constructed using only seismograms recorded at a single station from a
single event.  We identify patch space with the phase space,
reconstructed by time-delay embedding, of the dynamical system
associated with the earthquake.  In the following, we explore the
connections between the graph Laplacian defined on phase space and
several methods that have been proposed to characterize a dynamical
system.
%______________________________________________________________________________________________
\subsubsection{Recurrence quantification analysis}
%______________________________________________________________________________________________
We first observe that the weight matrix $\bW$ defined by
(\ref{weight}) is closely related to the recurrence plot matrix
associated with a dynamical system \citep{Eckmann87}. The recurrence
plot matrix $\bR$ is defined by
\begin{equation*}
  R_{i,j} = 
  \begin{cases}
    1 & \text{if $\|\bx_i - \bx_j\| < \delta$},\\
    0 & \text{otherwise}.
  \end{cases}
\end{equation*}
We can interpret the recurrence plot matrix $\bR$ in terms of the
weight matrix of a graph defined as follows: each vertex $\bx_i$ is
connected to the vertices that are within a distance $\delta$ of
$\bx_i$, and the weights along the edges are equal to one. This graph
is similar to the $\nu_{N\negmedspace N}$-nearest neighbor graph
used in this work when
$\delta$ is chosen so that on average each vertex is connected to
$\nu_{N\negmedspace N}$ vertices, and $\sigma= \infty$ in
(\ref{weight}). Several methods have been proposed to recover
information about $\bx(t)$ from the knowledge of $\bR$
\citep{Marwan07,Robinson09}. In this paper, we demonstrate that we can
indeed construct a smooth parametrization of the phase space
$\left\{\bx(t), t\ge 0\right\}$ from the knowledge of the weight
matrix $\bW$ only.
% ______________________________________________________________________________________________
\subsubsection{Complex networks in phase space}
%______________________________________________________________________________________________
The recurrence plot matrix $\bR$ can also be interpreted as the
adjacency matrix of a graph that captures the dynamical
system. Several authors have proposed to characterize nonlinear
dynamical systems by studying such graphs, known as ``recurrence
networks'' \cite{Donner10a,Gao09a,Marwan09} (see also
\cite{Shimada08,Xu08} for graphs that are constructed using the
$\nu_{N\negmedspace N}$ nearest neighbors in phase space, as we
do). After the graph is constructed, geometric properties of the
graph, such as diameter, path lengths distribution, are computed (see
\cite{Donner10a} and references therein for a detailed review).
Because such geometric properties can be computed directly from the
eigenvalues $\lambda_k$ defined by (\ref{geneigenprob})
\cite{Chung97}, the methods proposed in
\cite{Donner10a,Gao09a,Marwan09,Shimada08,Xu08} are in fact equivalent
to the problem of studying the Laplacian defined on the network. Our
approach can therefore be understood as a spectral characterization of
phase space obtained from the eigenfunctions and the eigenvalues of the
Laplacian defined on a recurrence network.
% ______________________________________________________________________________________________
\subsubsection{Frob\'enius-Perron Operator}
%______________________________________________________________________________________________
Finally, one can define a Frob\'enius-Perron operator, similar to the
matrix $\bP$, that characterizes the evolution of the probability
distribution of the configurations of phase space \cite{Collet06}. The
eigenfunctions of the Frob\'enius-Perron operator can be used to
partition phase space into almost invariant subsets
\cite{Dellnitz06,Dellnitz99,Froyland03}: a trajectory initiated in each
of these subsets remains in the subset for a long time before escaping
to another almost invariant subset \cite{Collet06}. The eigenfunctions
of the Perron-Frob\'enius operator are exactly similar to the $\bsi_k$. In
summary, in the simpler case of the analysis of a single earthquake
from a single station, our approach can be interpreted as a method to
decompose phase space in terms of regions where the long term dynamics
are similar \cite{Froyland03}.
%______________________________________________________________________________________________
\section{Estimation of Arrival-Times of Seismic Waves}
\label{detect}
% ______________________________________________________________________________________________
% ______________________________________________________________________________________________
\subsection{Learning the presence of seismic waves in patch-space} 
% ______________________________________________________________________________________________
Our goal is to learn the association between the presence of a seismic
wave within a patch, and the values of the patch coordinates.  As
explained before, we advocate a geometric approach: we expect that
patches will organize themselves on the unit sphere in $\R^{d-1}$ in a
manner that will reveal the presence of seismic waves.  We represent
all the patches with the coordinates defined by $\Bsi$ in
(\ref{embed}). We then use training data (labelled by experts) to
partially populate patch-space with information about the presence or
absence of seismic waves. We combine the information provided by the
labels with the knowledge about the geometry of patch-space to train a
classifier; this approach is known as semi-supervised
\cite{Chapelle06}. We then use the classifier to classify unlabelled
patches into baseline, or arrivals patches. The classification problem
is formulated as a kernel ridge regression problem \citep{Hastie09}:
for any given patch, the classifier returns a number between 0 and 1
that quantifies the probability that a seismic wave be present within
the patch.

We assume that $N_l$ of the $N$ patches have been labelled by an expert
(analyst): for each of these patches we know if a seismic wave was
observed in the patch, and at what time. We construct a response
function $f$ defined on the new coordinates, $\Bsi(\bx_i) \in \R^m$,
and taking values in $[0,1]$,
\begin{align}
f:  \R^m  \longrightarrow &[0,1]\\
\Bsi(\bx_i)  \longrightarrow & f(\Bsi(\bx_i)). 
\end{align}
The classifier decides that the patch $\bx_i$ contains an arrival if
the response $f(\Bsi(\bx_i))$ is greater than some threshold
$\varepsilon > 0$. The threshold $\varepsilon$ controls the rates of
false alarms and missed detections: a small $\varepsilon$ results in many
false alarms but will rarely miss arrivals, and vice versa. We expand
the response function as a linear combination of Gaussian kernels in
$\R^m$,
\begin{equation}
f (\Bsi(\bx)) =  \sum_{j = 1}^{N_l}\beta_j\,
\exp\left\{-\|\Bsi(\bx)-\Bsi(\bx_j)\|^2/\alpha^2\right\}. 
  \label{ridge}
\end{equation}
The vector of unknown coefficients $\bba = \begin{bmatrix} \beta_1,
  \ldots,\beta_{N_l} \end{bmatrix}$ is computed using the training
data. The kernel ridge regression \citep{Hastie09} combines two ideas:
distances between patches are measured using the Gaussian kernel
$\bK$, with entries $K_{i,j} = \exp
\left\{-\|\Bsi(\bx_j)-\Bsi(\bx_i)\|^2/\alpha^2\right\}$,
$i,j=1,\ldots, N_l$; and the classifier is designed to
provide the simplest model of the response in terms of the $N_l$
training data.  Rather than trying to find the optimal fit of the
function $f$ to the $N_l$ labelled patches, we penalize the regression
(\ref{ridge}) by imposing a penalty on the norm of $\bba$
\citep{Hastie09}. This prevents the model (\ref{ridge}) from overfitting the
training samples. The optimal regression is defined as the solution to
the quadratic minimization problem
\begin{equation}
\|\br - \bK \bba  \|^2 + \mu \bba^T\bK\bba,
\end{equation}
where $\br = \begin{bmatrix} r_1 & \ldots & r_{N_l}\end{bmatrix}^T$ is
the known response on the $N_l$ labelled patches. The parameter $\mu$ controls the amount of
penalization: $\mu=0$ yields a least squares fit, while $\mu = \infty$
ignores the data.  For a given choice of $\mu$, the optimal vector of
coefficients \citep{Hastie09} is given by
\begin{equation}
\boldsymbol\beta = (\bK+\mu \bI)^{-1}\br,
\label{beta}
\end{equation}
%
%______________________________________________________________________________________________
\begin{figure}[H]
  \centerline{
   \includegraphics[width =.8\textwidth]{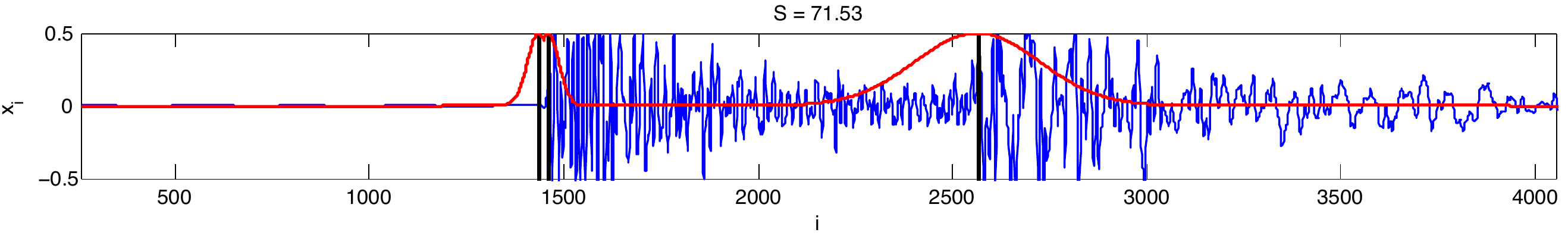}
}
  \centerline{
   \includegraphics[width =.8\textwidth]{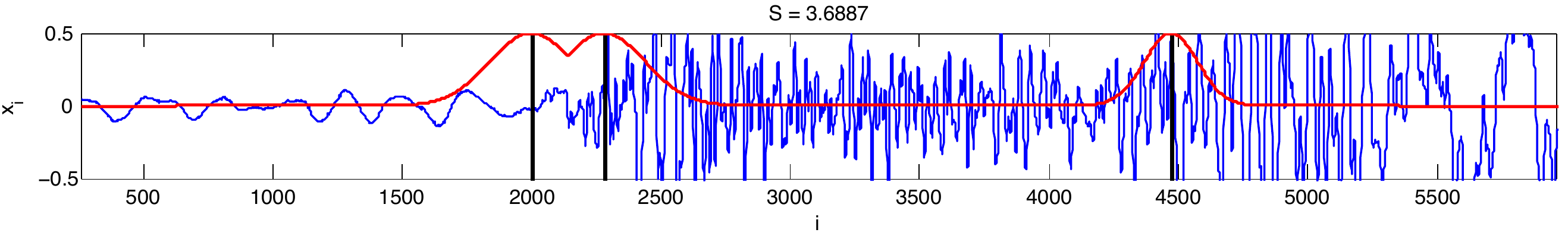}
 }
  \caption{Seismic traces $x_i$ (blue); estimated responses
    $r_i$ (red); arrival-times $\tau_J$ (black vertical bars).
    \label{h}}
\end{figure}
%______________________________________________________________________________________________
\noindent where $\bI$ is the $N_l \times N_l$ identity matrix. In our
experiments, the ridge parameter $\mu$ was determined by
cross-validation, and the same value, $\mu =0.8$, was used throughout.
The Gaussian width $\alpha$ is chosen to be a multiple of the average
kernel distance,
\begin{equation}
\alpha^2 = C \;\left (\text{average}_{\bx_i,\bx_j} \| \Bsi(\bx_i) - \Bsi(\bx_j)\|^2\right);
\label{alpha}
\end{equation}
for all experiments we chose $C= 0.51$.
%______________________________________________________________________________________________
\subsection{Defining  Ground truth}
\label{response}
% ______________________________________________________________________________________________
\subsubsection{Uncertainty in arrival-time}
%______________________________________________________________________________________________
In order to validate our approach we need to compare the output of the
response function $f$, defined by (\ref{ridge}), to the actual
decision provided by an expert (analyst). This comparison needs to be
performed for every patch being analyzed. Unfortunately, the decision
of the analyst is usually formulated as a binary response: an arrival
is present at time $\tau_i$ or not. We claim that this apparent
perfect determination of the arrival-time is misleading. Indeed, as
was pointed out in \citep{freedman66}, the origin time and the
arrival-time at a given station are, for all practical purposes,
random variables whose distributions depend on quality of the seismic
record and the training and experience of the analyst detecting the
arrivals.  We formalize this intuition and model the arrival-time
estimated by the analyst as a Gaussian distribution with mean $\tau_j$
and variance $h_j$. The parameter $h_j$ controls the width of the
Gaussian and quantifies the confidence with which the analyst
estimated $\tau_j$. Ideally, $h_j$ should be a function of the
inter-observer variability for the estimation of $\tau_j$. In this
work, we propose to estimate the uncertainty $h_j$ directly from the
seismogram. For each arrival-time $\tau_j$, we compute the dominant
frequency $\omega$ of $x(t)$ using a short Fourier transform. Let $T =
1/\omega$ be the period associated with $\omega$, we define the
uncertainty $h_j$ as follows
\begin{equation*}
h_j = 
\begin{cases}
2 T/\dt & \text{if $2T /\dt < h_{\max}$,}\\
h_{\max} & \text{otherwise.}
\end{cases}
\end{equation*}
This choice of $h_j$ corresponds to the following idea: if the
seismogram were to be a pure sinusoidal function oscillating at the
frequency $\omega$ (see Fig \ref{distance}), then this choice of $h_j$
would guarantee that we observe two periods (cycles) of $x(t)$ over a time
interval of length $h_j$.  

Finally, we define the true response $r_i$ at time $t_i$ to be 
the maximum of the Gaussian bumps associated with the arrival-time
times $\tau_j$ nearest to time $t_i$,
\begin{equation}
  r_i =  \max_j\left\{\exp\left(-{(t_i-\tau_j)^2}/h_j \right)\right\}.
\end{equation}
Figure \ref{h} displays two seismograms
with different values for $h_j$. In the top seismogram the first
arrival is very localized (small $h_1$), whereas the second arrival
corresponds to a lower instantaneous frequency, and is therefore less
localized (large $h_2$). In the second seismogram (bottom of
Fig.~\ref{h}) the first two arrivals are very close to one another
resulting in an overlap of the Gaussians defining the response $r_i$.
%
%______________________________________________________________________________________________
\begin{figure}[H]
{\SF A} \hspace*{-1pc} \centerline{
    \includegraphics[width =0.8\textwidth]{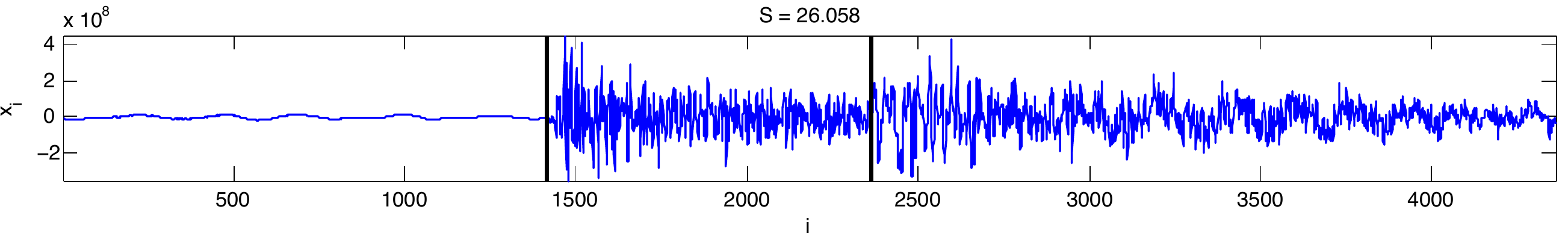}}
  \centerline{
    \includegraphics[width =0.8\textwidth]{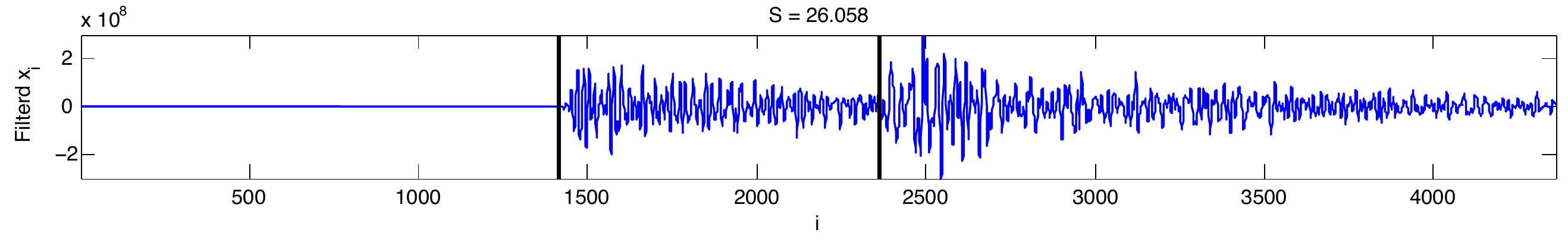}}
  \centerline{
    \includegraphics[width =0.8\textwidth]{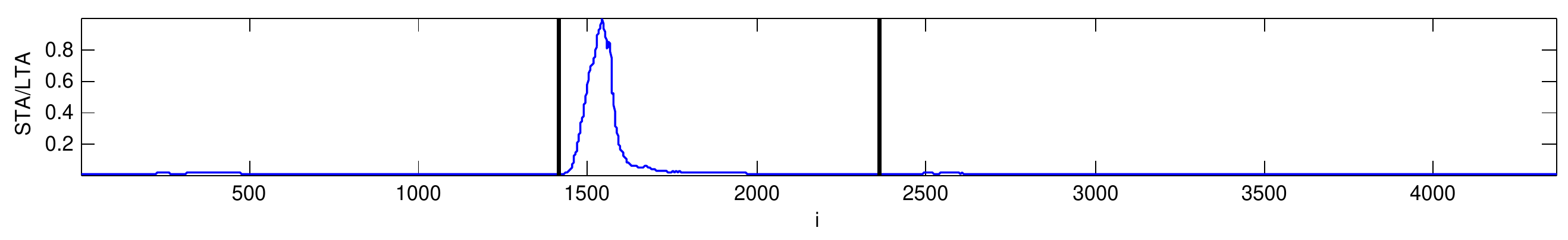}}
{\SF B} \hspace*{-3pc}\centerline{
    \includegraphics[width =0.8\textwidth]{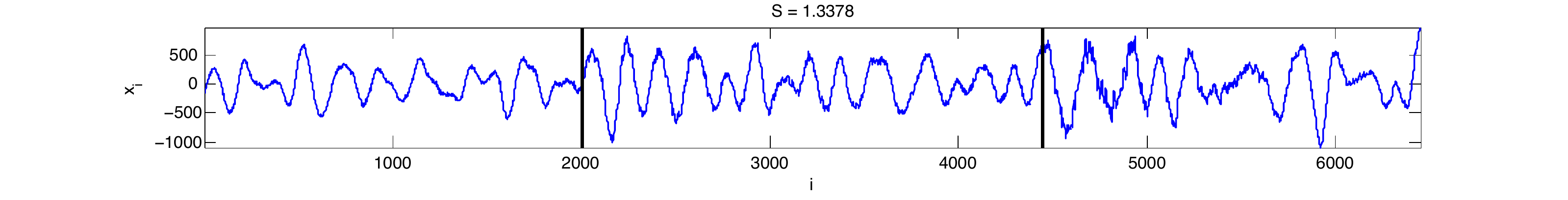}}
  \centerline{
    \includegraphics[width =0.8\textwidth]{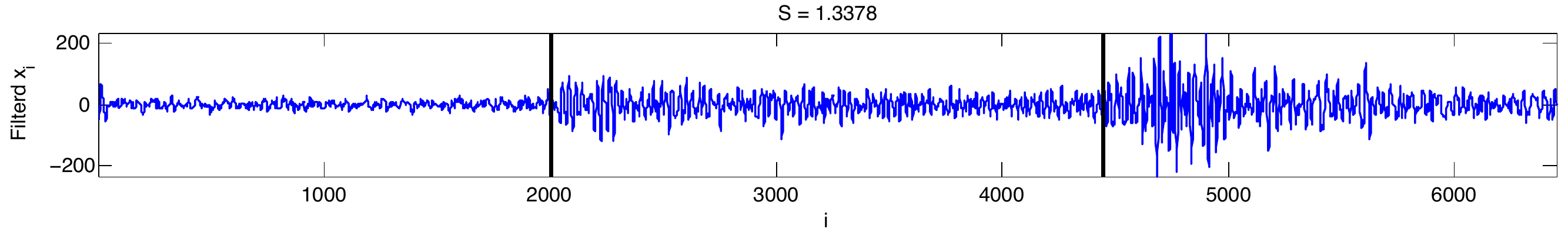}}
  \centerline{
    \includegraphics[width =0.8\textwidth]{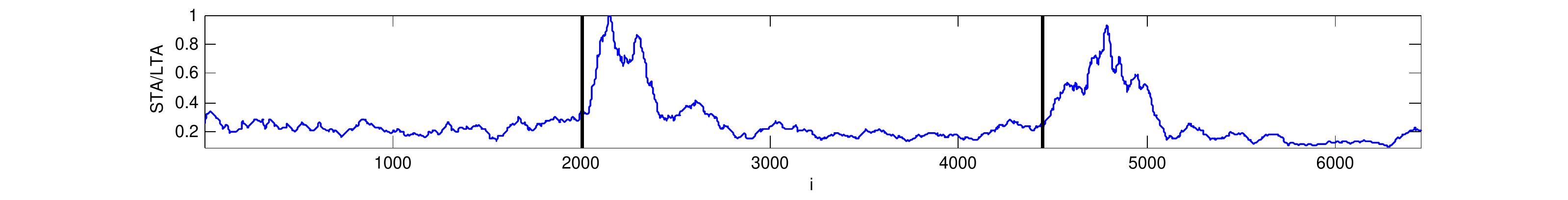}}
  \caption{Raw and filtered seismic traces with associated STA/LTA
    outputs;  high energy localization ({\SF
      A}) and very diffuse energy localization ({\SF B}).  Analyst picks are
    represented by bars.
    \label{differentS}}
\end{figure}
\noindent
%______________________________________________________________________________________________
\subsubsection{Energy localization of seismic traces}
%______________________________________________________________________________________________
Because we analyze seismic traces of very different quality, we need
to define a concept of energy localization associated with a seismic
trace. Indeed, variability in the estimation of arrival-times in human
is less pronounced when a seismic trace contains very localized
arrivals \citep{Velasco01}.  We plan therefore to evaluate the
performance of our method as a function of the energy localization
level. For this purpose, we define the energy localization of a given
trace to be the ratio of the energy of the seismic waves over the
energy of the baseline activity.  This ratio can be defined in
terms of the set of patches that are extracted from the trace. For a
given trace, let $\cal A$ be the subset of patches that contain
arrivals, and $\cal B$ the complement of $\cal A$: the subset of
patches that contain only baseline activity. We define the energy
localization by the ratio
\begin{equation}
  S = \frac{|{\cal B}|}{|{\cal A}|}
\left(    
  \sum_{\bx_i \in {\cal A}} \|\bx_i\|^2
\right) \left/
 \left(
    \sum_{\bx_i \in {\cal B}} \|\bx_i\|^2
  \right)\right. ,
  \label{figure:ratioofint}
\end{equation}
where $|\cal A|$ (resp. $|\cal B|$) is the cardinality of $\cal A$
(resp. $\cal B$).  Figure \ref{differentS} shows two seismic traces
with very different energy localizations.  Arrival-times assigned by
the analyst are represented by vertical bars.  The short window used
to compute the STA is three second long (120 samples) and the long
window, which immediately follows the short window, is 27 second long
(1080 samples).  Before computing the STA/LTA ratio, we apply a
standard preprocessing step and filter the raw seismic trace with a
passband ($[0.8,3.5]$ Hz)%
%______________________________________________________________________________________________
\begin{figure}[H]
  \centerline{
    \includegraphics[width = 12pc]{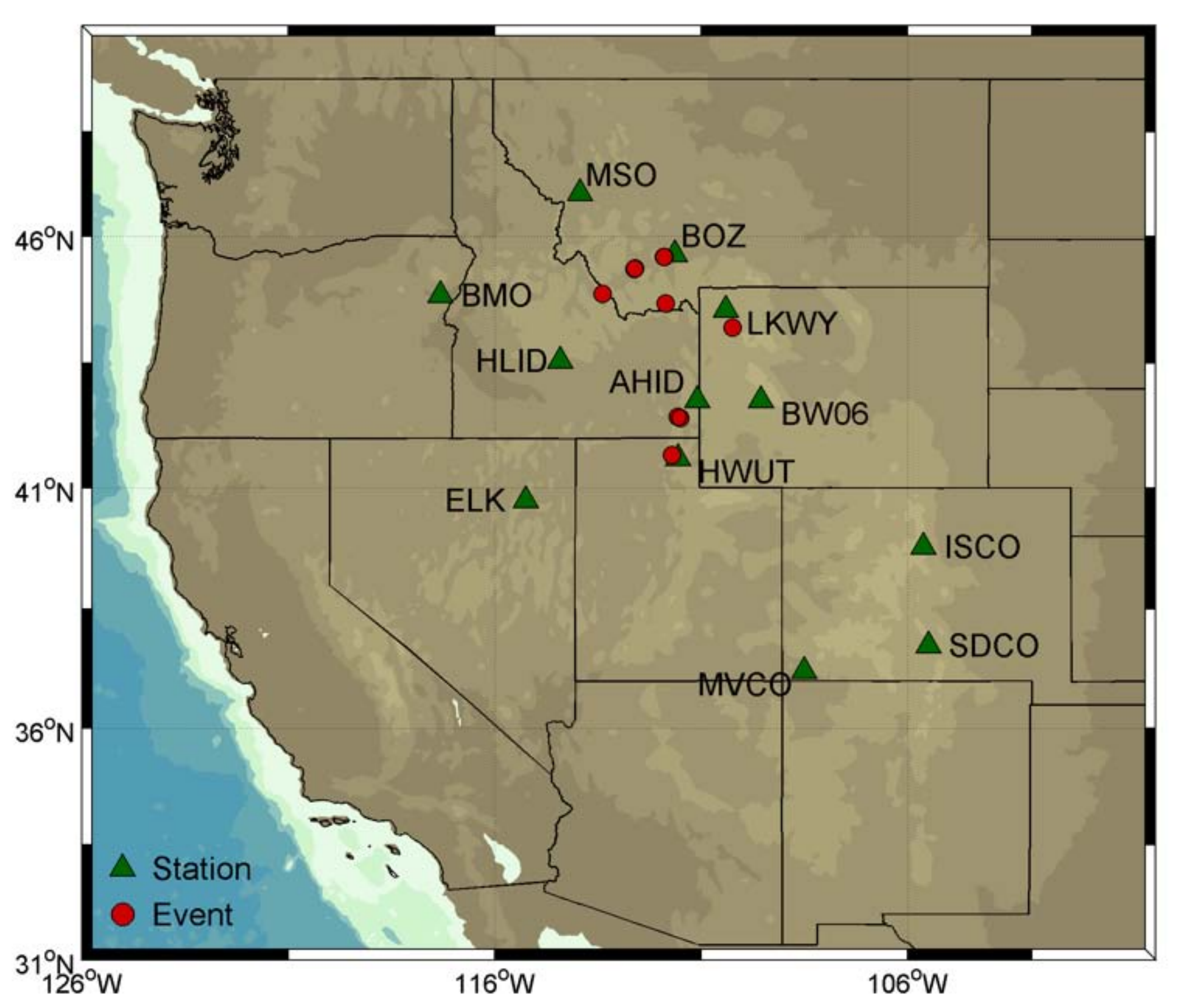}
  }
  \caption{Locations of the stations and events from the Rocky
    Mountain region. 
\label{seismicstation}}
\end{figure}
%______________________________________________________________________________________________ 
%
\noindent Butterworth filter (see
Fig.~\ref{differentS}). For the purpose of visual comparison, we
normalized the STA/LTA output so that its maximum value is one. The
trace ({\SF A}) in Fig.~\ref{differentS} has a large energy
localization, while the trace  ({\SF B}) has a very low energy
1localization.  The Butterworth filter is able to remove some of the
irrelevant low-frequency oscillations in ({\SF B}) and yields a signal
that can be processed by STA/LTA.  We note that the second arrival
(Lg) in the first trace ({\SF A}) is missed by the STA/LTA algorithm.
%______________________________________________________________________________________________
\section{Results}
\label{results}
% ______________________________________________________________________________________________
\subsection{Rocky Mountain Dataset}
\label{seisintro}
%_______________________________________________________________________________________________
We validate our approach with a dataset composed of broadband seismic
traces from seismic events that occurred in Idaho, Montana, Wyoming,
and Utah, between 2005 and 2006 (see
Fig.~\ref{seismicstation}). Arrival-times have been determined by an
analyst. The ten events with the largest number of arrivals were
selected for analysis.  In total, we used 84 different station records
from ten different events containing 226 labelled arrivals.  Of the
226 labelled arrivals, there are 72 Pn arrivals, 70 Pg arrivals, 6 Sn
arrivals, and 78 Lg arrivals. The sampling rate was $1/\dt = 40$ Hz.
We consider only the vertical channel in our analysis.  To minimize
the computational cost, patches are spaced apart by $40 \dt$ (one
second).
\subsection{Validation of the Classifier}
\label{validate}
%______________________________________________________________________________________________
The performance of the algorithm varies as a function of the energy
localization $S$, and therefore we perform three independent
validations by dividing the seismic traces into three homogeneous
subsets:$n_1=27$ traces with low energy localization ($S<3$), $n_2=29$
traces with medium energy localization ($3\leq S\leq18$), and the
remaining $n_3 =28$ traces with high energy localization (
$S>18$). For each subset $s, \; (s=1,2,3)$, we perform a standard
leave-one-out cross-validation \citep{Hastie09} using $n_s $ folds as
follows. We choose a test seismogram $x_t(t)$ among the $n_s$ traces
and compute the optimal set of weights (\ref{beta}) for the kernel
ridge classifier (\ref{ridge}) using the remaining $n_s -1$
traces. Patches $\bx_i$ are then randomly selected from the test
seismogram $x_t(t)$ and the classifier computes the response function
$f (\Bsi(\bx_i))$. The response of the classifier is compared to the
true response $r_i$ for various false alarms and missed detections
levels. We repeat this procedure for each possible test seismogram
$x_t(t)$ among the $n_s$ seismograms. Figure \ref{crossvalid} details
the cross-validation procedure. We quantify the performance of the
classifier using a Receiver Operating Characteristic (ROC) curve
\citep{Hastie09}. The true detection rate (one minus the type II
error) is plotted against the type I error (false alarm rate).  We
characterize each ROC curve by the area under the curve (the closer to
one, the better).
%______________________________________________________________________________________________
\subsection{Optimization of the parameters of the algorithm}
\begin{figure}[H]
  \centerline{
    \includegraphics[width = .8\textwidth]{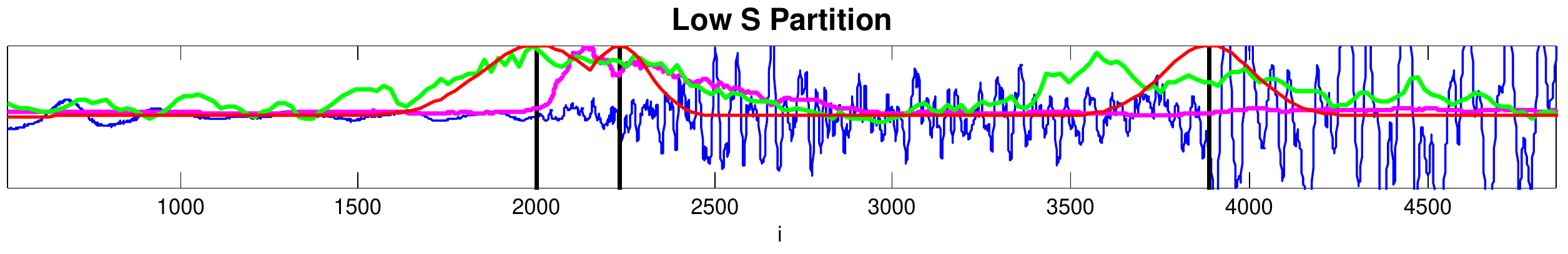}
}
  \centerline{
    \includegraphics[width = .8\textwidth]{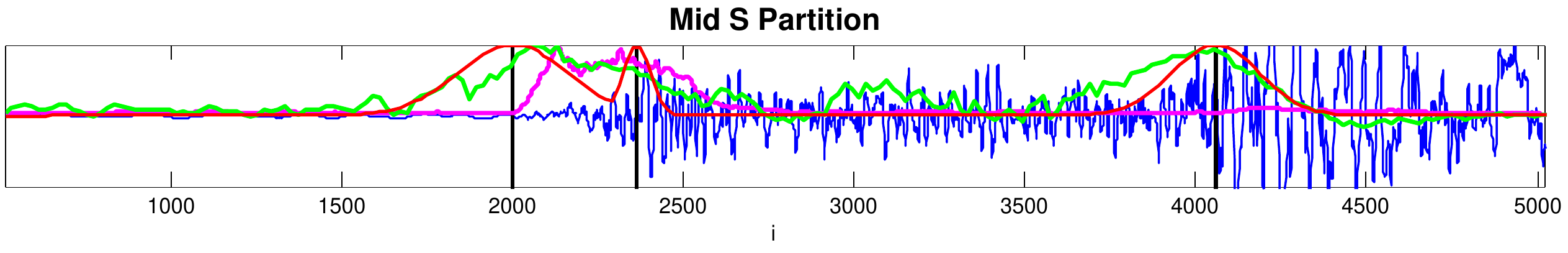}
}
  \centerline{
    \includegraphics[width = .8\textwidth]{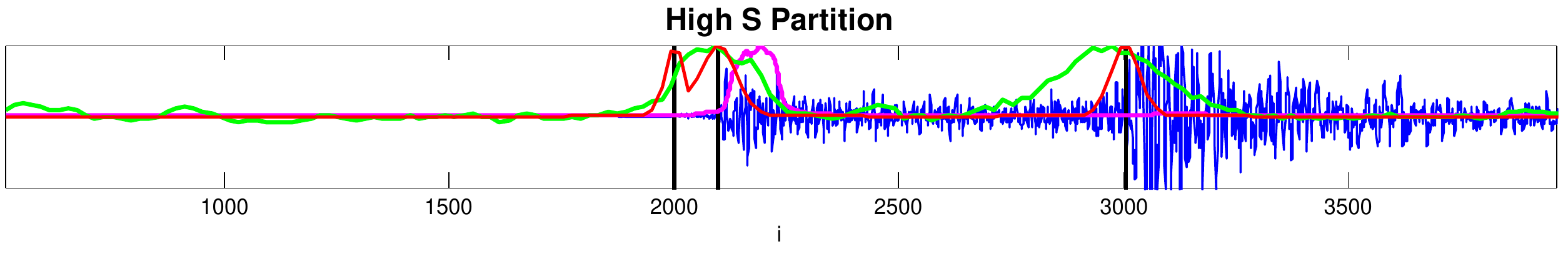}
}
\caption{Seismic trace $x_i$ (blue); true response $r_i$ (red);
  STA/LTA  (magenta); classifier $f(\Bsi(\bx_i))$
  (green).}
    \label{singletraceresponse}
\end{figure}

%______________________________________________________________________________________________
The optimal values of the parameters were computed using
cross-validation. This procedure turned out to be very robust, since
we used the same parameters for all experiments. The optimal
classification performance was achieved by choosing $\sigma = \infty$
and $\nu_{N\negmedspace N}=32$ in the construction of the graph
Laplacian. This is equivalent to setting the weights $W_{i,j}$ on the
edges to be 1, and yields a graph that is extremely robust to noise.
The influence of the patch size on the classification performance can
be found in a series of %
% __________________________________________________________________________________________
\begin{figure}[H]
  \makebox{\SF Algorithm 2: Cross validation of the classification}\\
  \barre
  \begin{itemize}
  \item [] {\SF Input}: Seismic traces, and the associated responses
    $(r_i)$. 
  \item []{\SF Algorithm:}
  \item [] {\SF for} $s=1$ {\SF to} $3$ \hfill // {\it for each subset of seismic traces}
    \begin{enumerate}
    \item [] extract a total of $N$ patches from $n_s$ distinct seismic traces.
    \item [] compute new coordinates $\Bsi(\bx_i)$ of each patch $\bx_i$ $i=1,\ldots,N$
    \item [] {\SF for} $j=1$ {\SF to} $n_s$ \hfill // {\it evaluate the classifier
        for each seismic trace j} 
      \begin{enumerate}
      \item [] build classifier using all patches except those from
        trace $j$
      \item [] {\SF for all} patches $\bx^j_i$ in trace $j$ compute
        classifier response $f(\Bsi(\bx^j_i))$
      \item [] {\SF for} $\varepsilon= \varepsilon_l$ {\SF to} $\varepsilon_u$ 
      \item [] // {\it populate the ROC curve using different
          thresholds to detect an arrival}  
        \begin{itemize}
        \item [] {\SF if} $f(\Bsi(\bx^j_i))>\varepsilon$ {\SF and}
          $r_i<\varepsilon_0$ {\SF then} declare false positive
       \item [] {\SF else if} $f(\Bsi(\bx^j_i))<\varepsilon$ {\SF and}
          $r_i>\varepsilon_0$ {\SF then} declare false negative
       \item []{\SF end if}
        \item [] record false positive and false negative
          rates for patches in fold $j$
        \end{itemize}
      \item [] {\SF end for}
      \end{enumerate}
    \item [] {\SF end for}
    \item [] compute average false positive and false negative
      rate
    \end{enumerate}
 \item [] {\SF end for}
  \item [] {\SF Output}: area under the ROC curve.
  \end{itemize}
\caption{Cross validation procedure. \label{crossvalid}}
\end{figure}
\noindent  ROC curves in Fig.~\ref{roc}. We observe in
Fig.~\ref{roc} that the STA/LTA algorithm performs best for
seismograms with low energy localization. Indeed, waveforms with high
energy localization contain very little energy in the baseline part of
the waveform. As a result, the STA/LTA ratio is much larger for the
primary waves than it is for the secondary waves, and STA/LTA often
misses the secondary waves. For traces with low energy localization,
the ratio between baseline and arrivals energies is much smaller. This
causes the STA/LTA ratio to reach similar values for the primary and
secondary arrivals. Figure \ref{singletraceresponse} shows three
seismic traces with different energy localizations. STA/LTA (magenta)
misses the secondary wave for medium and high energy localizations.
Our approach (green) can detect the primary and secondary waves at all
energy localization levels. For seismic traces with low energy
localization our approach cannot compete with STA/LTA when $d$ drops
below 256. Of course, it is unfair to compare our approach using
patches of only 128 samples with STA/LTA, which uses 1080 time samples
(27 seconds). Indeed, as soon as $d \ge 1024$, our approach
outperforms STA/LTA. Interestingly, our approach does not benefit from
using a much larger $d$; when $d=2048$
the scale of the local analysis is no longer adapted to the physical
process that we study.
%_________________________________________________________________________________
\subsection{So what does the set of patches look like?}
\begin{figure}[H]
\centerline{
  \includegraphics[width = .3\textwidth]{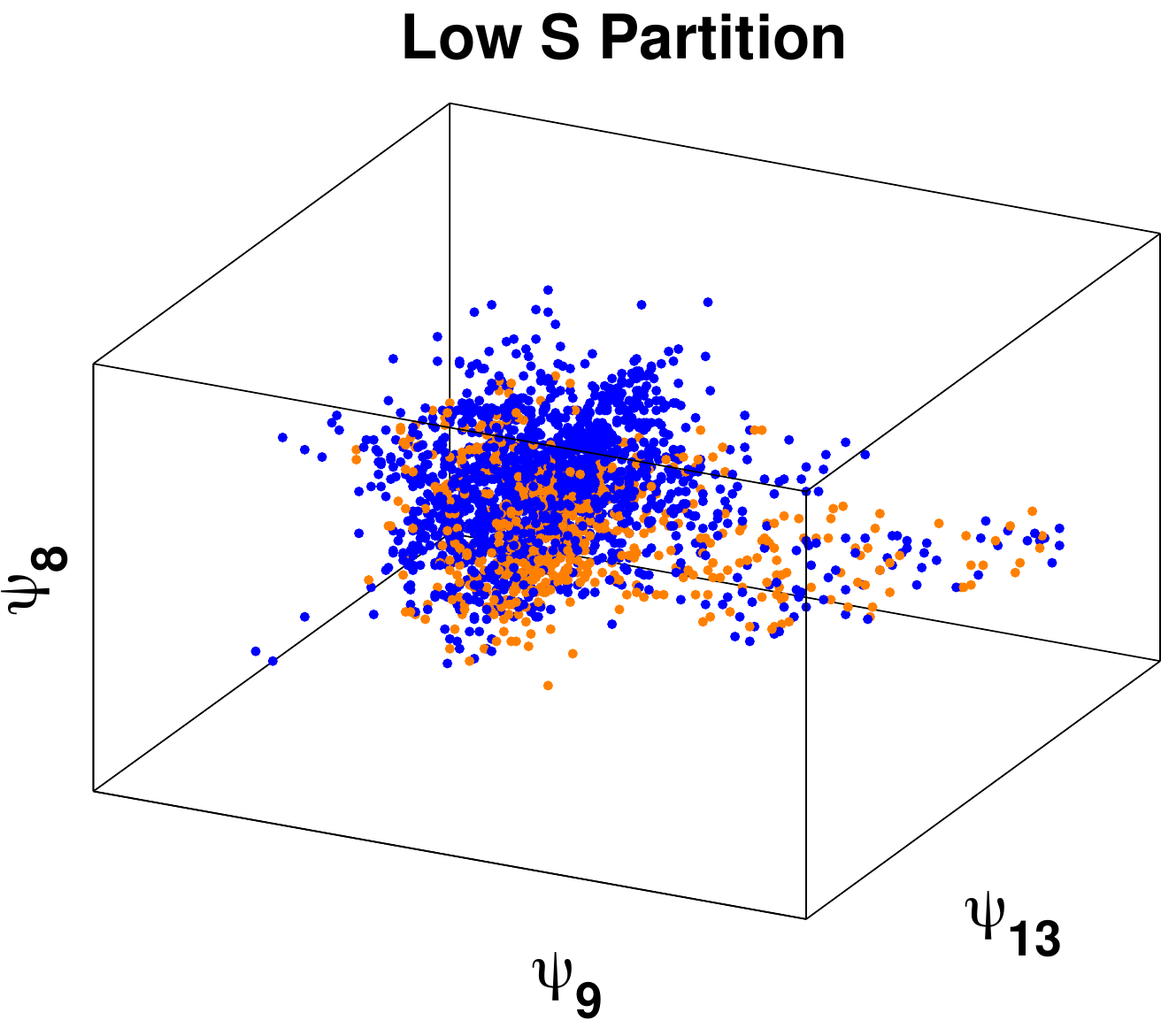}
  \includegraphics[width = .3\textwidth]{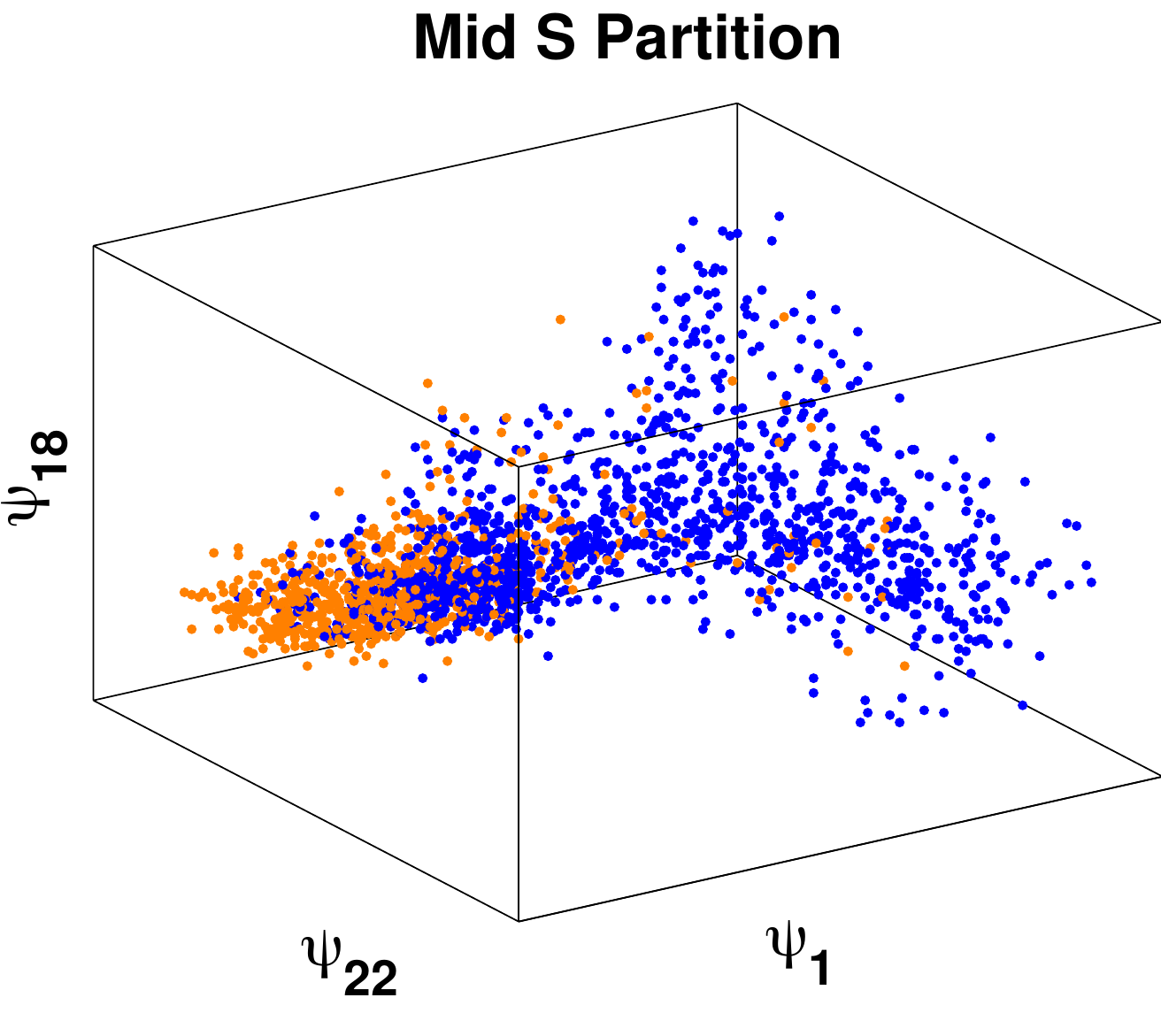}
  \includegraphics[width = .3\textwidth]{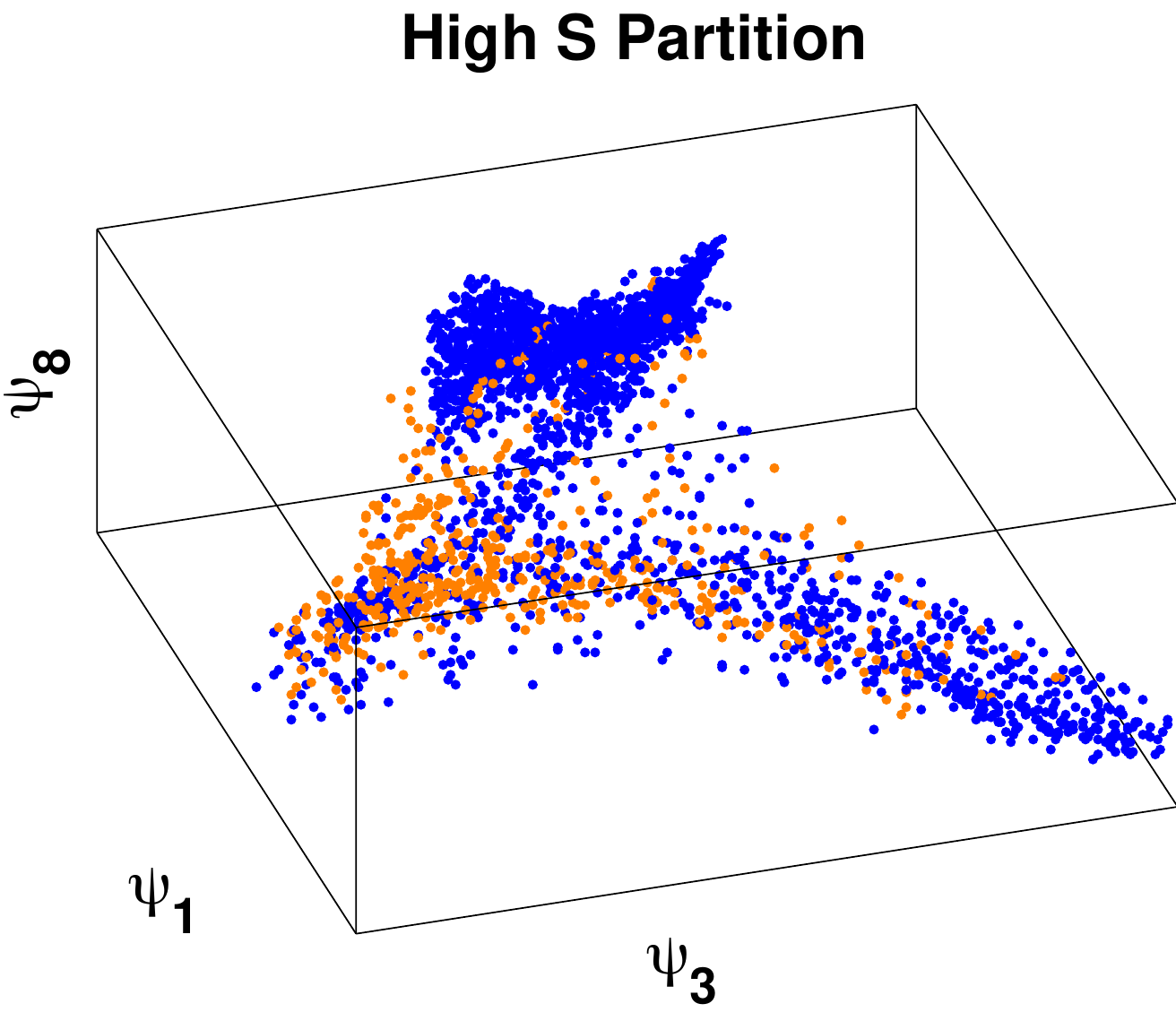}
}
\caption{Each patch $\bx_i$ is represented as a dot using three
  coordinates of the vector $\Bsi$.  The color encodes the presence
  (orange) or absence (blue) of an arrival within $\bx_i$. The energy
  localization levels increases from left to right.}
\label{seismicembed}
\end{figure}

%_________________________________________________________________________________
To help us gain some insight into the geometric organization of patch
space we display the patches using some of the new coordinates $\Bsi
(\bx_i)$.  For the three subsets of patches (classified according to
the energy localization), we display in Fig.~\ref{seismicembed} each
patch as a dot using three coordinates of the coordinate vectors%
$\Bsi$. Because we can only display three coordinates, among the 25 (or
50) that we use to classify the patches, we use the three coordinates that
best reveal the organization of patch-space.  The color of the dot
encodes the presence (orange) or absence (blue) of an arrival within
$\bx_i$.  As the energy localization increases the separation between
baseline patches and arrival patches improves. This visual impression
is confirmed using the quantitative evaluation performed with the ROC
curves (see Fig. \ref{roc}). Clearly the shape of the set of patches
is not linear, and would not be well approximated with a linear
subspace.
%______________________________________________________________________________________________
%______________________________________________________________________________________________
\noindent
%______________________________________________________________________________________________
\subsection{Classification Performance}
\label{benchmark}
%______________________________________________________________________________________________
We first compare our approach to the gold-standard provided by
STA/LTA. The second stage of the evaluation consists in quantifying
the importance of the nonlinear dimension reduction $\Bsi$ defined by
(\ref{embed}). To gauge the effect of $\Bsi$ we replace it by two
linear transforms: a wavelet transform and a PCA transform. In both
cases, we reduce the dimensionality of each patch from $d$ to $m$.
Wavelets have been used for a long time in seismology because
seismograms can be approximated with very high precision using a small
number of wavelet coefficients (e.g. \citep{Anant97,Zhang03,Gendron00}
and references therein). On the other hand, we can also try to find
the best linear approximation to a set of $N$ patches. This linear
approximation is obtained using PCA (also known as the
singular-spectrum analysis \citep{Vautard89}, see section
\ref{discrete}). The first $m$ vectors of a PCA analysis yields the
subspace that provides the optimal $m$-dimensional approximation to
the set of patches. Our experiments demonstrate that the set of
patches is not a linear structure and therefore is poorly approximated
using PCA.
%______________________________________________________________________________________________
\subsubsection{STA/LTA Ratio}
%______________________________________________________________________________________________
We implement an STA/LTA detector as follows. We first apply a passband
([0.8~-~3.5] Hz) Butterworth filter to the raw traces. We then compute
the ratio of the energy over two adjacent time window: a short window
of 120 time samples (2 s) immediately followed by a long window of
1080 time samples (27 s). An arrival is detected when this ratio
exceeds a threshold.
%______________________________________________________________________________________________
%______________________________________________________________________________________________
\subsubsection{PCA and Wavelet Representations of Patch-Space}
%______________________________________________________________________________________________
An orthonormal wavelet transform (symmlet 8) provides a multiscale
decomposition of each patch $\bx_i$ in terms of $d$ coefficients. Many
of the coefficients are small and can be ignored.  In order to decide
which wavelet coefficients to retain, we select a fixed set of $m/2$
indices corresponding to the largest coefficients of the baseline
patches. Similarly, we select the $m/2$ indices associated to the
largest coefficients among the patches that contain arrivals. This
procedure allows us to define a fixed set of $m$ wavelet coefficients
that are used for all patches as in input to the ridge regression
algorithm. Similarly, we keep the first $m$ coordinates returned by
PCA.
%______________________________________________________________________________________________
\subsubsection{Parameters of the classifier based on the PCA and Wavelet Representations}
%______________________________________________________________________________________________
After applying a wavelet transform, or PCA, we use the same ridge
classifier to detect arrivals.  The parameters of the classifier are
optimized for the wavelet and PCA transforms respectively.  The
Gaussian width $\alpha$ was again chosen to be a multiple of the
average kernel distance between any two patches (see
(\ref{alpha})). The parameter $C$ in equation (\ref{alpha}) was set to
$C= 6.9$ for the wavelet parametrization, and $C=4.6$ for the PCA
parametrization. The ridge regression parameter was the same for both
wavelets and PCA and was equal to $\mu = 10^{-3}$.
%
%__________________________________________________________________________________________
\begin{figure}[H]
\centerline{
  \includegraphics[width = .33\textwidth]{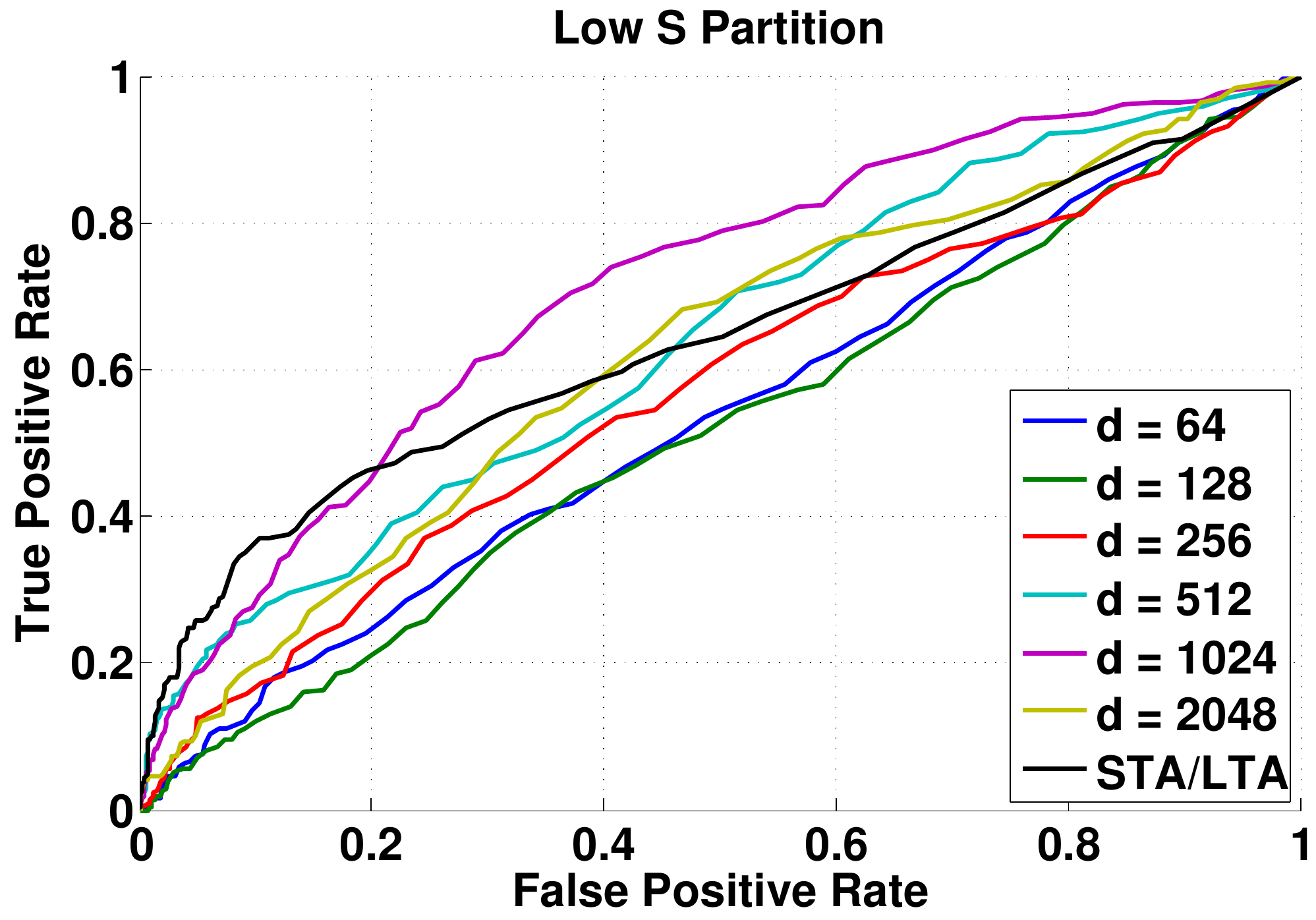}
  \includegraphics[width = .33\textwidth]{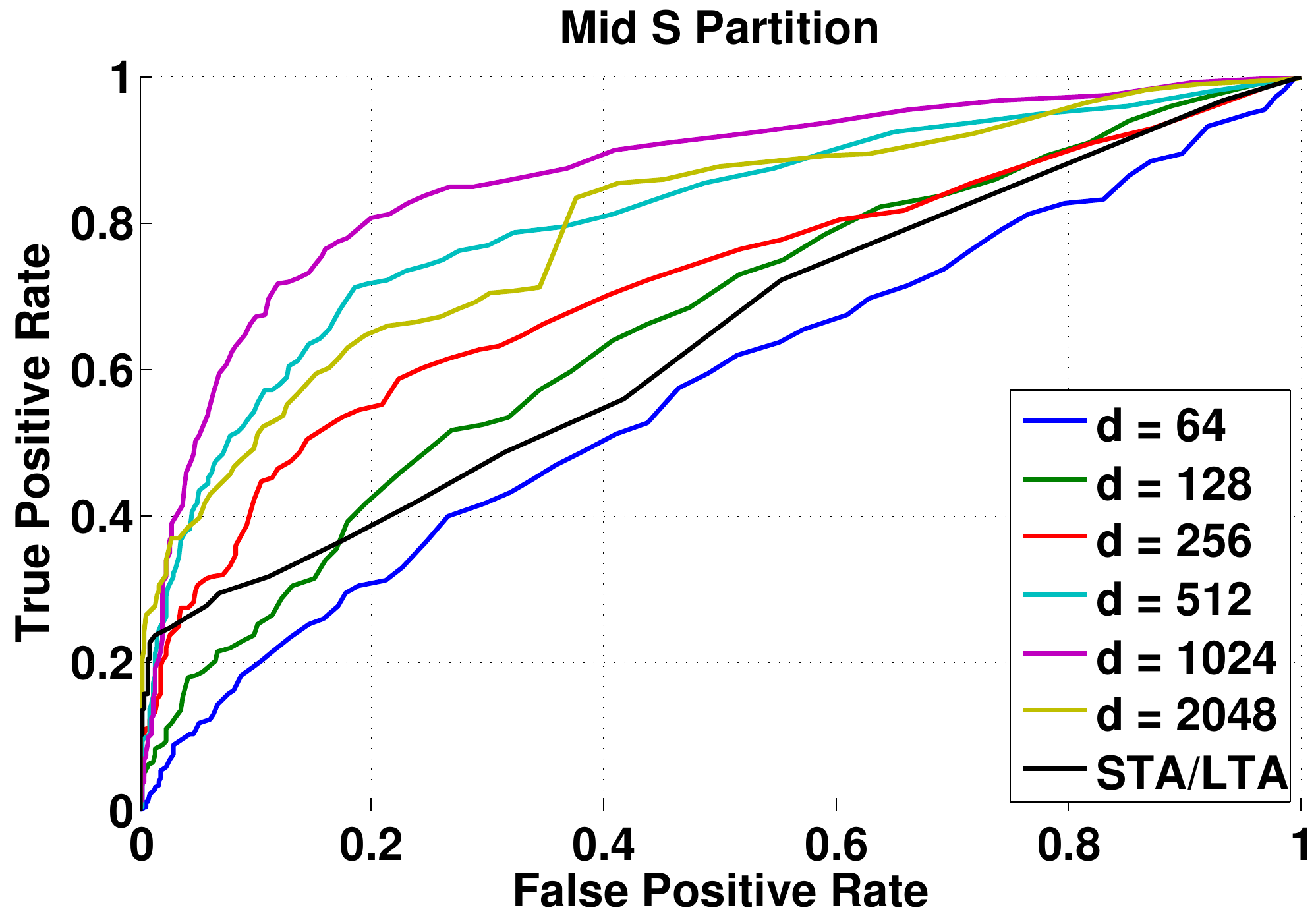}
  \includegraphics[width = .33\textwidth]{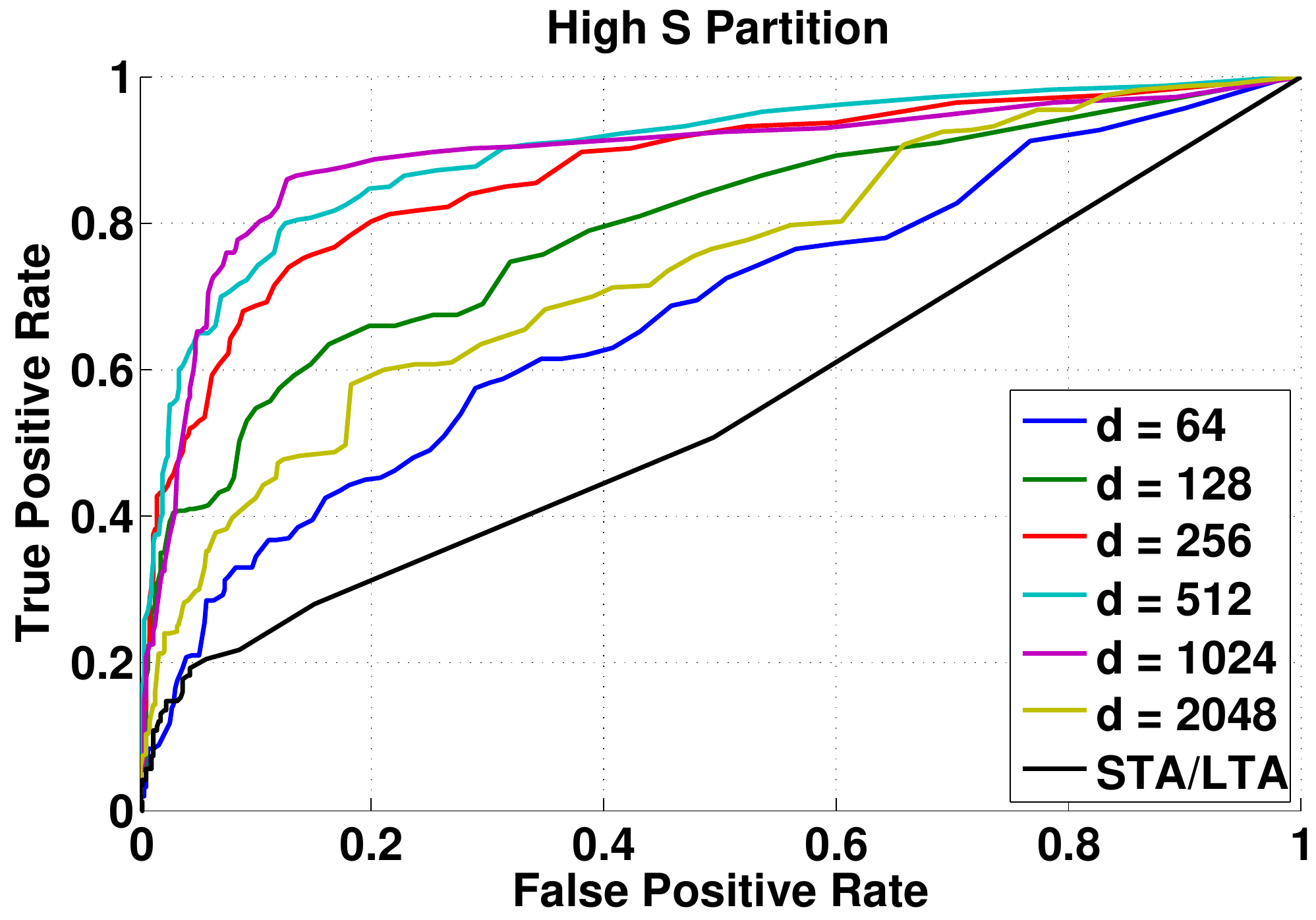}
}
    \caption{ROC curves for various values of the embedding dimension $d$
      at three levels of energy localization.} 
    \label{roc}
\end{figure}%
\noindent 
%__________________________________________________________________________________________
\noindent
%
%______________________________________________________________________________________________
\section{Discussion}
%______________________________________________________________________________________________
Table \ref{tableroc} provides a detailed summary of the performance of
our approach. For each energy localization level (see section
\ref{validate} for the definition of the subsets), we report the
performance of the different detection methods as a function of $d$
(patch dimension) and $m$ (reduced dimension). The performance is
quantified using the area under the ROC curve (the ROC curves are
shown in Fig.~\ref{roc}); a perfect detector should have an area equal to one.\\

{\noindent \bf Effect of the patch size.} As expected, the patch-based
methods perform poorly if the patch is too small (there is not enough
information to detect the seismic wave) or too large (the information
is smeared over too large a window). The choice of the optimal patch
size is dictated by the physical processes at stake here, since the
optimal size is the same for all methods, irrespective of the
transform used to reduce dimensionality. For high energy localization
seismograms, the seismic waves are very localized and therefore all
algorithms perform better with smaller
patches (256 or 512 instead of 1024).\\

{\noindent \bf Effect of the transform used to reduce dimensionality.}
The experiments indicate that PCA outperforms a wavelet decomposition
at every energy localization level. Both PCA and the wavelet transform
are orthonormal transforms that can be understood in terms of a
rotation of patch-space. PCA provides the optimal rotation to align
patch-space along the $m$-dimensional subspace of best-fit.  Finally,
the nonlinear transformation $\Bsi$ based on the eigenfunctions of the
Laplacian outperforms both PCA and wavelets. This%
%
%______________________________________________________________________________________________
\begin{table}[H]
\caption{Area under the ROC
  curve (closer to 1 is better) as a function of  the
  patch dimension $d$  and the reduced dimension $m$, at three
  different energy localization levels $S$.
\label{tableroc}}
\centerline{
\begin{tabular}{@{}lrlllllll@{}} \toprule%
&d & STA/LTA & \multicolumn{2}{c}{Wavelet}  & \multicolumn{2}{c}{PCA} & \multicolumn{2}{c}{Laplacian} \\
dimension &       &     & 25 & 50 & 25 & 50 & 25 & 50\\\hline
Low $S$ 
& 64    & --  &  0.53 & \red 0.53 & 0.51 & 0.55 &0.53 & 0.53\\
& 128   & --  & \red 0.55 & 0.49 & 0.52 & 0.51 &0.52 & 0.52\\
& 256   & --  &  0.52 & 0.47 & 0.51 & 0.54 &0.54 & 0.57\\
& 512   & --  &   0.53 &  0.50 &  \red 0.61 &  0.61 &0.62 & 0.64\\
& 1024  & 0.66&  0.43 & 0.38 & 0.54 & \red 0.64 &\red  0.70 & \red  0.71\\
& 2048  & --  &  0.45 & 0.39 & 0.55 & 0.48 &0.61 & 0.62\\\hline
Mid $S$                                     
& 64    & --  &  0.54 & 0.52 & 0.52 & 0.54 &0.57 & 0.57\\
& 128   & --  &  0.57 & 0.55 & 0.53 & 0.56 &0.66 & 0.66\\
& 256   & --  &  0.61 & 0.62 & 0.70 & 0.71 &0.71 & 0.71\\
& 512   & --  &  0.68 & 0.67 & 0.76 & 0.79 &0.79 & 0.81\\
& 1024  & 0.68&  \red  0.77 & \red  0.76 &  \red 0.81 &  \red 0.84 &\red  0.86 &  \red 0.86\\
& 2048  & --  &  0.64 & 0.66 & 0.69 & 0.75 &0.80 & 0.80\\\hline
High $S$                                    
& 64    & --  &  0.56 & 0.62 & 0.65 & 0.65 & 0.72 & 0.67\\
& 128   & --  &  0.68 & 0.69 &  \red 0.78 & 0.73 &0.80 & 0.79\\
& 256   & --  &  0.72 & 0.70 & 0.77 &  0.84 & 0.88 &  0.87\\
& 512   & --  &   \red 0.74 & \red  0.79 &  0.73 &  \red 0.85 &  \red 0.90 &\red   0.90\\
& 1024  & 0.59&  0.72 & 0.76 & 0.67 & 0.75 & 0.88 & 0.89\\
& 2048  & --  &  0.51 & 0.49 & 0.57 & 0.67 &0.76 & 0.74\\
\bottomrule
\end{tabular}
}
\end{table}
%______________________________________________________________________________________________
\noindent clearly indicates
that the set of patches contains nonlinear structures that cannot be
well approximated by the optimal linear subspace computed by
PCA. Interestingly, the results (not shown) are not improved by
applying a wavelet transform before applying the nonlinear map $\Bsi$
(\ref{embed}) (see \cite{Schclar10} for an example of a combination of
wavelet transform with a nonlinear map similar to $\Bsi$).\\

{\noindent \bf Dimension of patch-space.} The performance is not
significantly improved when 50 coordinates are used instead of
25. This is a result that is independent of the method used to reduced
dimensionality, and is therefore a statement about the complexity of
patch-space and about the physical nature of the seismic traces.  As
mentioned before, several studies have estimated the dimensionality of
the low-dimensional inertial manifold reconstructed from the phase
space of the tremors of a single volcano. This dimensionality was
found in most studies \citep{Delauro08,Demartino04,Konstantinou02} to
be less than five: a number much smaller than our rough estimate of
the dimensionality of patch space.  Because patch space includes
several seismograms from different events measured at different
stations, we expect the dimensionality of this set to be greater than
the dimensionality of the phase space reconstructed from the tremors
of a single volcano measured at a single station. On the other hand,
our study confirms that the combined phase spaces associated with
regional seismic waves remains remarkably low-dimensional.

{\noindent \bf Computational complexity} The complexity of this method
is mainly determined by the combined complexity of the nearest
neighbor search and the eigenvalue problem.  We currently use the
restarted Arnoldi method for sparse matrices implemented by the Matlab
function {\tt eigs} to solve the eigenvalue problem.  We use the fast
approximate nearest neighbor algorithm implemented by the ANN library
\cite{Arya98} to construct the graph of patches.\\

{\noindent \bf Parametrization of slow manifolds with the eigenvectors
  of the Laplacian.} As discussed in Sec. \ref{discrete}, our approach
is related to the problem of parametrizing the low dimensional
attracting manifold associated with the nonlinear dynamical system
that is at the origin of the seismic waves. Similar ideas have been
proposed in the context of molecular dynamics. In \cite{Coifman08},
the authors estimate slow variables that can be used to coarse-grain
the dynamics near an inertial manifold. The slow variables are the
eigenfunctions of a Fokker-Planck diffusion process defined on
simulated data.  Similarly, it was shown in
\cite{Deuflhard00,Dellnitz06} that invariant subsets associated with a
dynamical system can be identified by studying the eigenvectors of a
probability transition matrix -- similar to the row-stochastic matrix
$\bD^{-1}\bW$ defined in (\ref{geneigenprob}) -- in the phase space of
the dynamical system. Horenko \cite{Horenko08} proposes to reconstruct
the phase space generated by time-delay embedding using a combination
of $K$ of $m$-dimensional subspaces. In contrast, we use a nonlinear
parametrization in this paper.

\section{Concluding remarks}
In this paper we presented a novel method to estimate from a
seismogram arrival-times of seismic waves. We use time-delay embedding
to characterize the local dynamics over temporal patches extracted
from the seismic waveforms. We combine several delay-coordinates
phase spaces formed by the different patch trajectories, and construct
a graph that quantifies the distances between the different temporal
patches. The eigenvectors of the Laplacian defined on the graph
provide a low-dimensional parametrization of the combined
phase spaces.  Finally, a kernel ridge regression learns the
association between each configuration of the phase space and the
presence of a seimic wave. The regression is performed using the
low-dimensional parametrization of the set of patches.  Our approach
outperforms standard linear techniques, such as wavelets and
singular-spectrum analysis, and makes it possible to capture the
nonlinear structures of the phase space reconstructed from time-delay
embedding. This method unites the existing theory on time-delay
embedding and the recent results on the nonlinear parametrization of
manifold-valued data \cite{Belkin03,Coifman06b,Coifman08,Jones08}. We
expect that our approach may be applicable to time series that are
generated by other complex nonlinear dynamical processes, such as
neurophysiological data, financial data, etc. We also expect that the
idea of reconstructing phase space using the eigenvectors of the graph
Laplacian may be used to remove noise from data generated by nonlinear
dynamical systems \cite{Broomhead96,Kostelich93}, and predict
time-series \cite{Casdagli89,Kantz04}.
%______________________________________________________________________________________________
\subsection*{Acknowledgments}
This work was supported by National Nuclear Security Administration
Contract No. DE-AC04-94AL8500.%

\subsection*{Author Contributions}
Conception and design of methods and experiments: FGM and KMT.
Acquisition of data: MJP and CJY. Analysis and interpretation of data: KMT,
MJP, CJY, and FGM. Writing of manuscript: FGM and KMT.%
%\bibliographystyle{model1-num-names}
%\bibliography{/Users/francois/LaTeX/Bib/biblio}

\begin{thebibliography}{67}
\expandafter\ifx\csname natexlab\endcsname\relax\def\natexlab#1{#1}\fi
\providecommand{\bibinfo}[2]{#2}
\ifx\xfnm\relax \def\xfnm[#1]{\unskip,\space#1}\fi
%Type = Article
\bibitem[{Allen(1982)}]{Allen82}
\bibinfo{author}{R.~Allen},
\newblock \bibinfo{title}{Automatic phase pickers: Their present use and future
  prospects},
\newblock \bibinfo{journal}{Bull. seism. Soc. Am.} \bibinfo{volume}{68}
  (\bibinfo{year}{1982}) \bibinfo{pages}{1521--1532}.
%Type = Article
\bibitem[{Panagiotakis et~al.(2008)Panagiotakis, Kokinou, and
  Vallianatos}]{Panagiotakis08}
\bibinfo{author}{C.~Panagiotakis}, \bibinfo{author}{E.~Kokinou},
  \bibinfo{author}{F.~Vallianatos},
\newblock \bibinfo{title}{{Automatic $ P $-Phase Picking Based on Local-Maxima
  Distribution}},
\newblock \bibinfo{journal}{IEEE Trans. Geosci. Remote Sens.}
  \bibinfo{volume}{46} (\bibinfo{year}{2008}) \bibinfo{pages}{2280--2287}.
%Type = Article
\bibitem[{Di~Stefano et~al.(2006)Di~Stefano, Aldersons, Kissling, Baccheschi,
  Chiarabba, and Giardini}]{Distefano06}
\bibinfo{author}{R.~Di~Stefano}, \bibinfo{author}{F.~Aldersons},
  \bibinfo{author}{E.~Kissling}, \bibinfo{author}{P.~Baccheschi},
  \bibinfo{author}{C.~Chiarabba}, \bibinfo{author}{D.~Giardini},
\newblock \bibinfo{title}{{Automatic seismic phase picking and consistent
  observation error assessment: application to the Italian seismicity}},
\newblock \bibinfo{journal}{Geophys. J. Intern.} \bibinfo{volume}{165}
  (\bibinfo{year}{2006}) \bibinfo{pages}{121--134}.
%Type = Article
\bibitem[{Freiberger(1963)}]{Freiberger63}
\bibinfo{author}{W.~Freiberger},
\newblock \bibinfo{title}{{An approximate method in signal detection}},
\newblock \bibinfo{journal}{Quarterly App. Math} \bibinfo{volume}{20}
  (\bibinfo{year}{1963}) \bibinfo{pages}{373--378}.
%Type = Techreport
\bibitem[{Berger and Sax(2001)}]{Berger01}
\bibinfo{author}{J.~Berger}, \bibinfo{author}{R.~Sax}, \bibinfo{title}{Seismic
  detectors: the state of the art}, \bibinfo{type}{Technical Report}, VELA
  Seismological Center, Alexandria, VA, \bibinfo{year}{2001}.
%Type = Article
\bibitem[{Persson(2003)}]{Persson03}
\bibinfo{author}{L.~Persson},
\newblock \bibinfo{title}{{Statistical tests for regional seismic phase
  characterizations}},
\newblock \bibinfo{journal}{J. Seismol.} \bibinfo{volume}{7}
  (\bibinfo{year}{2003}) \bibinfo{pages}{19--33}.
%Type = Article
\bibitem[{Saragiotis et~al.(2002)Saragiotis, Hadjileontiadis, and
  Panas}]{Saragiotis02}
\bibinfo{author}{C.~Saragiotis}, \bibinfo{author}{L.~Hadjileontiadis},
  \bibinfo{author}{S.~Panas},
\newblock \bibinfo{title}{{PAI-S/K: A robust automatic seismic P phase arrival
  identification scheme}},
\newblock \bibinfo{journal}{IEEE Trans. Geosci. Remote Sens.}
  \bibinfo{volume}{40} (\bibinfo{year}{2002}) \bibinfo{pages}{1395--1404}.
%Type = Article
\bibitem[{K{\"u}perkoch et~al.(2010)K{\"u}perkoch, Meier, Lee, Friederich
  et~al.}]{Kuperkoch10}
\bibinfo{author}{L.~K{\"u}perkoch}, \bibinfo{author}{T.~Meier},
  \bibinfo{author}{J.~Lee}, \bibinfo{author}{W.~Friederich}, et~al.,
\newblock \bibinfo{title}{{Automated determination of P-phase arrival times at
  regional and local distances using higher order statistics}},
\newblock \bibinfo{journal}{Geophys. J. Intern.} \bibinfo{volume}{181}
  (\bibinfo{year}{2010}) \bibinfo{pages}{1159--1170}.
%Type = Article
\bibitem[{Galiana-Merino et~al.(2008)Galiana-Merino, Rosa-Herranz, and
  Parolai}]{Galiana02}
\bibinfo{author}{J.~Galiana-Merino}, \bibinfo{author}{J.~Rosa-Herranz},
  \bibinfo{author}{S.~Parolai},
\newblock \bibinfo{title}{{Seismic $P$ Phase Picking Using a Kurtosis-Based
  Criterion in the Stationary Wavelet Domain}},
\newblock \bibinfo{journal}{IEEE Trans. Geosci. Remote Sens.}
  \bibinfo{volume}{46} (\bibinfo{year}{2008}) \bibinfo{pages}{3815--3826}.
%Type = Article
\bibitem[{Zhang et~al.(2003)Zhang, Thurber, and Rowe}]{Zhang03}
\bibinfo{author}{H.~Zhang}, \bibinfo{author}{C.~Thurber},
  \bibinfo{author}{C.~Rowe},
\newblock \bibinfo{title}{{Automatic P-wave arrival detection and picking with
  multiscale wavelet analysis for single-component recordings}},
\newblock \bibinfo{journal}{Bull. seism. Soc. Am.} \bibinfo{volume}{93}
  (\bibinfo{year}{2003}) \bibinfo{pages}{1904}.
%Type = Article
\bibitem[{Bardainne et~al.(2006)Bardainne, Gaillot, Dubos-Sall{\'e}e, Blanco,
  and S\'en\'echal}]{Bardainne06}
\bibinfo{author}{T.~Bardainne}, \bibinfo{author}{P.~Gaillot},
  \bibinfo{author}{N.~Dubos-Sall{\'e}e}, \bibinfo{author}{J.~Blanco},
  \bibinfo{author}{G.~S\'en\'echal},
\newblock \bibinfo{title}{Characterization of seismic waveforms and
  classification of seismic events using chirplet atomic decomposition},
\newblock \bibinfo{journal}{Geophys. J. Intern.} \bibinfo{volume}{166}
  (\bibinfo{year}{2006}) \bibinfo{pages}{699--718}.
%Type = Article
\bibitem[{Withers et~al.(1998)Withers, Aster, Young, Beiriger, Harris, Moore,
  and Trujillo}]{Withers98}
\bibinfo{author}{M.~Withers}, \bibinfo{author}{R.~Aster},
  \bibinfo{author}{C.~Young}, \bibinfo{author}{J.~Beiriger},
  \bibinfo{author}{M.~Harris}, \bibinfo{author}{S.~Moore},
  \bibinfo{author}{J.~Trujillo},
\newblock \bibinfo{title}{A comparison of select trigger algorithms for
  automated global seismic phase and event detection},
\newblock \bibinfo{journal}{Bull. seism. Soc. Am.} \bibinfo{volume}{88(1)}
  (\bibinfo{year}{1998}) \bibinfo{pages}{95--106}.
%Type = Article
\bibitem[{Wang(2002)}]{Wang02}
\bibinfo{author}{J.~Wang},
\newblock \bibinfo{title}{Adaptive training of neural networks for automatic
  seismic phase identification},
\newblock \bibinfo{journal}{Pure appl. Geophys.} \bibinfo{volume}{159}
  (\bibinfo{year}{2002}) \bibinfo{pages}{1021--1041}.
%Type = Article
\bibitem[{Sauer et~al.(1991)Sauer, Yorke, and Casdagli}]{Sauer91}
\bibinfo{author}{T.~Sauer}, \bibinfo{author}{J.~Yorke},
  \bibinfo{author}{M.~Casdagli},
\newblock \bibinfo{title}{{Embedology}},
\newblock \bibinfo{journal}{J. Stat. Phys.} \bibinfo{volume}{65}
  (\bibinfo{year}{1991}) \bibinfo{pages}{579--616}.
%Type = Article
\bibitem[{Abarbanel et~al.(1993)Abarbanel, Brown, Sidorowich, and
  Tsimring}]{Abarbanel93}
\bibinfo{author}{H.~Abarbanel}, \bibinfo{author}{R.~Brown},
  \bibinfo{author}{J.~Sidorowich}, \bibinfo{author}{L.~Tsimring},
\newblock \bibinfo{title}{{The analysis of observed chaotic data in physical
  systems}},
\newblock \bibinfo{journal}{Rev. Modern Phys.} \bibinfo{volume}{65}
  (\bibinfo{year}{1993}) \bibinfo{pages}{1331--1392}.
%Type = Article
\bibitem[{Gilmore(1998)}]{Gilmore98}
\bibinfo{author}{R.~Gilmore},
\newblock \bibinfo{title}{{Topological analysis of chaotic dynamical systems}},
\newblock \bibinfo{journal}{Rev. Modern Phys.} \bibinfo{volume}{70}
  (\bibinfo{year}{1998}) \bibinfo{pages}{1455--1529}.
%Type = Article
\bibitem[{Packard et~al.(1980)Packard, Crutchfield, Farmer, and
  Shaw}]{Packard80}
\bibinfo{author}{N.~Packard}, \bibinfo{author}{J.~Crutchfield},
  \bibinfo{author}{J.~Farmer}, \bibinfo{author}{R.~Shaw},
\newblock \bibinfo{title}{{Geometry from a time series}},
\newblock \bibinfo{journal}{Phys. Rev. Lett.} \bibinfo{volume}{45}
  (\bibinfo{year}{1980}) \bibinfo{pages}{712--716}.
%Type = Inproceedings
\bibitem[{Aster and Rowe(2000)}]{Aster00}
\bibinfo{author}{R.~Aster}, \bibinfo{author}{C.~Rowe},
\newblock \bibinfo{title}{Automatic phase pick refinement and similar event
  association in large seismic datasets},
\newblock in: \bibinfo{booktitle}{Advances in seismic event location},
  volume~\bibinfo{volume}{18}, pp. \bibinfo{pages}{231--263}.
%Type = Article
\bibitem[{Ben-Zion(2008)}]{BenZion08}
\bibinfo{author}{Y.~Ben-Zion},
\newblock \bibinfo{title}{Collective behavior of earthquakes and faults:
  Continuum-discrete transitions, progressive evolutionary changes, and
  different dynamic regimes},
\newblock \bibinfo{journal}{Rev. Geophys.} \bibinfo{volume}{46}
  (\bibinfo{year}{2008}).
%Type = Article
\bibitem[{Judd and Mees(1998)}]{Judd98}
\bibinfo{author}{K.~Judd}, \bibinfo{author}{A.~Mees},
\newblock \bibinfo{title}{{Embedding as a modeling problem}},
\newblock \bibinfo{journal}{Physica D} \bibinfo{volume}{120}
  (\bibinfo{year}{1998}) \bibinfo{pages}{273--286}.
%Type = Article
\bibitem[{Kennel and Abarbanel(2002)}]{Kennel02}
\bibinfo{author}{M.~Kennel}, \bibinfo{author}{H.~Abarbanel},
\newblock \bibinfo{title}{{False neighbors and false strands: A reliable
  minimum embedding dimension algorithm}},
\newblock \bibinfo{journal}{Phys. Rev. E} \bibinfo{volume}{66}
  (\bibinfo{year}{2002}) \bibinfo{pages}{26209}.
%Type = Article
\bibitem[{Small and Tse(2004)}]{Small04}
\bibinfo{author}{M.~Small}, \bibinfo{author}{C.~Tse},
\newblock \bibinfo{title}{{Optimal embedding parameters: a modelling
  paradigm}},
\newblock \bibinfo{journal}{Physica D} \bibinfo{volume}{194}
  (\bibinfo{year}{2004}) \bibinfo{pages}{283--296}.
%Type = Book
\bibitem[{Scott(1992)}]{Scott92}
\bibinfo{author}{D.~Scott}, \bibinfo{title}{Multivariate Density Estimation},
  \bibinfo{publisher}{Wiley}, \bibinfo{year}{1992}.
%Type = Article
\bibitem[{Broomhead and King(1986)}]{Broomhead86}
\bibinfo{author}{D.~Broomhead}, \bibinfo{author}{G.~King},
\newblock \bibinfo{title}{{Extracting qualitative dynamics from experimental
  data}},
\newblock \bibinfo{journal}{Physica D} \bibinfo{volume}{20}
  (\bibinfo{year}{1986}) \bibinfo{pages}{217--236}.
%Type = Article
\bibitem[{Vautard and Ghil(1989)}]{Vautard89}
\bibinfo{author}{R.~Vautard}, \bibinfo{author}{M.~Ghil},
\newblock \bibinfo{title}{{Singular spectrum analysis in nonlinear dynamics,
  with applications to paleoclimatic time series}},
\newblock \bibinfo{journal}{Physica D} \bibinfo{volume}{35}
  (\bibinfo{year}{1989}) \bibinfo{pages}{395--424}.
%Type = Article
\bibitem[{Sharma et~al.(1993)Sharma, Vassiliadis, and Papadopoulos}]{Sharma93}
\bibinfo{author}{A.~Sharma}, \bibinfo{author}{D.~Vassiliadis},
  \bibinfo{author}{K.~Papadopoulos},
\newblock \bibinfo{title}{{Reconstruction of low-dimensional magnetospheric
  dynamics by singular spectrum analysis}},
\newblock \bibinfo{journal}{Geophys. Res. Lett.} \bibinfo{volume}{20}
  (\bibinfo{year}{1993}) \bibinfo{pages}{335--338}.
%Type = Article
\bibitem[{G{\'a}miz-Fortis et~al.(2002)G{\'a}miz-Fortis, Pozo-V{\'a}zquez,
  Esteban-Parra, and Castro-D{\'\i}ez}]{Gamiz02}
\bibinfo{author}{S.~G{\'a}miz-Fortis}, \bibinfo{author}{D.~Pozo-V{\'a}zquez},
  \bibinfo{author}{M.~Esteban-Parra}, \bibinfo{author}{Y.~Castro-D{\'\i}ez},
\newblock \bibinfo{title}{{Spectral characteristics and predictability of the
  NAO assessed through Singular Spectral Analysis}},
\newblock \bibinfo{journal}{J. Geophys. Res} \bibinfo{volume}{107}
  (\bibinfo{year}{2002}) \bibinfo{pages}{4685}.
%Type = Article
\bibitem[{De~Lauro et~al.(2008)De~Lauro, De~Martino, Del~Pezzo, Falanga, Palo,
  and Scarpa}]{Delauro08}
\bibinfo{author}{E.~De~Lauro}, \bibinfo{author}{S.~De~Martino},
  \bibinfo{author}{E.~Del~Pezzo}, \bibinfo{author}{M.~Falanga},
  \bibinfo{author}{M.~Palo}, \bibinfo{author}{R.~Scarpa},
\newblock \bibinfo{title}{{Model for high-frequency Strombolian tremor inferred
  by wavefield decomposition and reconstruction of asymptotic dynamics}},
\newblock \bibinfo{journal}{J. Geophys. Res.} \bibinfo{volume}{113}
  (\bibinfo{year}{2008}) \bibinfo{pages}{B02302}.
%Type = Article
\bibitem[{De~Martino et~al.(2004)De~Martino, Falanga, and Godano}]{Demartino04}
\bibinfo{author}{S.~De~Martino}, \bibinfo{author}{M.~Falanga},
  \bibinfo{author}{C.~Godano},
\newblock \bibinfo{title}{{Dynamical similarity of explosions at Stromboli
  volcano}},
\newblock \bibinfo{journal}{Geophys. J. Intern.} \bibinfo{volume}{157}
  (\bibinfo{year}{2004}) \bibinfo{pages}{1247--1254}.
%Type = Article
\bibitem[{Chouet and Shaw(1991)}]{Chouet91}
\bibinfo{author}{B.~Chouet}, \bibinfo{author}{H.~Shaw},
\newblock \bibinfo{title}{{Fractal properties of tremor and gas piston events
  observed at Kilauea Volcano Hawaii}},
\newblock \bibinfo{journal}{J. Geophys. Res} \bibinfo{volume}{96}
  (\bibinfo{year}{1991}) \bibinfo{pages}{10177--10189}.
%Type = Article
\bibitem[{Godano et~al.(1996)Godano, Cardaci, and Privitera}]{Godano96}
\bibinfo{author}{C.~Godano}, \bibinfo{author}{C.~Cardaci},
  \bibinfo{author}{E.~Privitera},
\newblock \bibinfo{title}{{Intermittent behaviour of volcanic tremor at Mt.
  Etna}},
\newblock \bibinfo{journal}{Pure appl. Geophys.} \bibinfo{volume}{147}
  (\bibinfo{year}{1996}) \bibinfo{pages}{729--744}.
%Type = Article
\bibitem[{Yuan et~al.(2004)Yuan, Lozier, Pratt, Jones, and Helfrich}]{Yuan04}
\bibinfo{author}{G.~Yuan}, \bibinfo{author}{M.~Lozier},
  \bibinfo{author}{L.~Pratt}, \bibinfo{author}{C.~Jones},
  \bibinfo{author}{K.~Helfrich},
\newblock \bibinfo{title}{{Estimating the predictability of an oceanic time
  series using linear and nonlinear methods}},
\newblock \bibinfo{journal}{J. Geophys. Res} \bibinfo{volume}{109}
  (\bibinfo{year}{2004}).
%Type = Article
\bibitem[{Konstantinou and Schlindwein(2002)}]{Konstantinou02b}
\bibinfo{author}{K.~Konstantinou}, \bibinfo{author}{V.~Schlindwein},
\newblock \bibinfo{title}{{Nature, wavefield properties and source mechanism of
  volcanic tremor: a review}},
\newblock \bibinfo{journal}{J. Volcanol. Geoth. Res.} \bibinfo{volume}{119}
  (\bibinfo{year}{2002}) \bibinfo{pages}{161--187}.
%Type = Article
\bibitem[{Konstantinou(2002)}]{Konstantinou02}
\bibinfo{author}{K.~Konstantinou},
\newblock \bibinfo{title}{{Deterministic non-linear source processes of
  volcanic tremor signals accompanying the 1996 Vatnajokull eruption, central
  Iceland}},
\newblock \bibinfo{journal}{Geophys. J. Intern.} \bibinfo{volume}{148}
  (\bibinfo{year}{2002}) \bibinfo{pages}{663--675}.
%Type = Book
\bibitem[{Chung(1997)}]{Chung97}
\bibinfo{author}{F.~Chung}, \bibinfo{title}{Spectral Graph Theory},
  \bibinfo{publisher}{CBNS-AMS}, \bibinfo{year}{1997}.
%Type = Article
\bibitem[{B\'erard et~al.(1994)B\'erard, Besson, and Gallot}]{Berard94}
\bibinfo{author}{P.~B\'erard}, \bibinfo{author}{G.~Besson},
  \bibinfo{author}{S.~Gallot},
\newblock \bibinfo{title}{Embeddings \mbox{Riemannian} manifolds by their heat
  kernel},
\newblock \bibinfo{journal}{Geom. Funct. Anal.} \bibinfo{volume}{4}
  (\bibinfo{year}{1994}) \bibinfo{pages}{373--398}.
%Type = Article
\bibitem[{Belkin and Niyogi(2003)}]{Belkin03}
\bibinfo{author}{M.~Belkin}, \bibinfo{author}{P.~Niyogi},
\newblock \bibinfo{title}{Laplacian eigenmaps for dimensionality reduction and
  data representation},
\newblock \bibinfo{journal}{Neural Comput.} \bibinfo{volume}{15}
  (\bibinfo{year}{2003}) \bibinfo{pages}{1373--1396}.
%Type = Article
\bibitem[{Coifman and Maggioni(2006)}]{Coifman06b}
\bibinfo{author}{R.~R. Coifman}, \bibinfo{author}{M.~Maggioni},
\newblock \bibinfo{title}{Diffusion wavelets},
\newblock \bibinfo{journal}{Appl. Comput. Harmon. A.} \bibinfo{volume}{21}
  (\bibinfo{year}{2006}) \bibinfo{pages}{53 -- 94}.
%Type = Article
\bibitem[{Coifman et~al.(2008)Coifman, Kevrekidis, Lafon, Maggioni, and
  Nadler}]{Coifman08}
\bibinfo{author}{R.~R. Coifman}, \bibinfo{author}{I.~G. Kevrekidis},
  \bibinfo{author}{S.~Lafon}, \bibinfo{author}{M.~Maggioni},
  \bibinfo{author}{B.~Nadler},
\newblock \bibinfo{title}{Diffusion maps, reduction coordinates, and low
  dimensional representation of stochastic systems},
\newblock \bibinfo{journal}{Multiscale Model. Sim.} \bibinfo{volume}{7}
  (\bibinfo{year}{2008}) \bibinfo{pages}{842--864}.
%Type = Article
\bibitem[{Jones et~al.(2008)Jones, Maggioni, and Schul}]{Jones08}
\bibinfo{author}{P.~W. Jones}, \bibinfo{author}{M.~Maggioni},
  \bibinfo{author}{R.~Schul},
\newblock \bibinfo{title}{{Manifold parametrizations by eigenfunctions of the
  Laplacian and heat kernels}},
\newblock \bibinfo{journal}{P. Natl. Acad. Sci. USA} \bibinfo{volume}{105}
  (\bibinfo{year}{2008}) \bibinfo{pages}{1803--1808}.
%Type = Article
\bibitem[{Saito(2008)}]{Saito08}
\bibinfo{author}{N.~Saito},
\newblock \bibinfo{title}{Data analysis and representation on a general domain
  using eigenfunctions of laplacian},
\newblock \bibinfo{journal}{Appl. Comput. Harmon. A.} \bibinfo{volume}{25}
  (\bibinfo{year}{2008}) \bibinfo{pages}{68--97}.
%Type = Article
\bibitem[{Eckmann et~al.(1987)Eckmann, Oliffson~Kamphorst, and
  Ruelle}]{Eckmann87}
\bibinfo{author}{J.~Eckmann}, \bibinfo{author}{S.~Oliffson~Kamphorst},
  \bibinfo{author}{D.~Ruelle},
\newblock \bibinfo{title}{{Recurrence Plots of Dynamical Systems}},
\newblock \bibinfo{journal}{Europhys. Lett.} \bibinfo{volume}{4}
  (\bibinfo{year}{1987}) \bibinfo{pages}{973--977}.
%Type = Article
\bibitem[{Marwan et~al.(2007)Marwan, Carmen~Romano, Thiel, and
  Kurths}]{Marwan07}
\bibinfo{author}{N.~Marwan}, \bibinfo{author}{M.~Carmen~Romano},
  \bibinfo{author}{M.~Thiel}, \bibinfo{author}{J.~Kurths},
\newblock \bibinfo{title}{{Recurrence plots for the analysis of complex
  systems}},
\newblock \bibinfo{journal}{Phys. Rep.} \bibinfo{volume}{438}
  (\bibinfo{year}{2007}) \bibinfo{pages}{237--329}.
%Type = Article
\bibitem[{Robinson and Thiel(2009)}]{Robinson09}
\bibinfo{author}{G.~Robinson}, \bibinfo{author}{M.~Thiel},
\newblock \bibinfo{title}{{Recurrences determine the dynamics}},
\newblock \bibinfo{journal}{Chaos} \bibinfo{volume}{19} (\bibinfo{year}{2009})
  \bibinfo{pages}{023104}.
%Type = Article
\bibitem[{Donner et~al.(2010)Donner, Zou, Donges, Marwan, and
  Kurths}]{Donner10a}
\bibinfo{author}{R.~Donner}, \bibinfo{author}{Y.~Zou},
  \bibinfo{author}{J.~Donges}, \bibinfo{author}{N.~Marwan},
  \bibinfo{author}{J.~Kurths},
\newblock \bibinfo{title}{{Recurrence networks—a novel paradigm for nonlinear
  time series analysis}},
\newblock \bibinfo{journal}{New J. Phys.} \bibinfo{volume}{12}
  (\bibinfo{year}{2010}) \bibinfo{pages}{033025}.
%Type = Article
\bibitem[{Gao and Jin(2009)}]{Gao09a}
\bibinfo{author}{Z.~Gao}, \bibinfo{author}{N.~Jin},
\newblock \bibinfo{title}{{Complex network from time series based on phase
  space reconstruction}},
\newblock \bibinfo{journal}{Chaos} \bibinfo{volume}{19} (\bibinfo{year}{2009})
  \bibinfo{pages}{033137}.
%Type = Article
\bibitem[{Marwan et~al.(2009)Marwan, Donges, Zou, Donner, and
  Kurths}]{Marwan09}
\bibinfo{author}{N.~Marwan}, \bibinfo{author}{J.~Donges},
  \bibinfo{author}{Y.~Zou}, \bibinfo{author}{R.~Donner},
  \bibinfo{author}{J.~Kurths},
\newblock \bibinfo{title}{{Complex network approach for recurrence analysis of
  time series}},
\newblock \bibinfo{journal}{Phys. Lett. A} \bibinfo{volume}{373}
  (\bibinfo{year}{2009}) \bibinfo{pages}{4246--4254}.
%Type = Article
\bibitem[{Shimada et~al.(2008)Shimada, Kimura, and Ikeguchi}]{Shimada08}
\bibinfo{author}{Y.~Shimada}, \bibinfo{author}{T.~Kimura},
  \bibinfo{author}{T.~Ikeguchi},
\newblock \bibinfo{title}{{Analysis of Chaotic Dynamics Using Measures of the
  Complex Network Theory}},
\newblock \bibinfo{journal}{Artificial Neural Networks-ICANN 2008}
  (\bibinfo{year}{2008}) \bibinfo{pages}{61--70}.
%Type = Article
\bibitem[{Xu et~al.(2008)Xu, Zhang, and Small}]{Xu08}
\bibinfo{author}{X.~Xu}, \bibinfo{author}{J.~Zhang},
  \bibinfo{author}{M.~Small},
\newblock \bibinfo{title}{{Superfamily phenomena and motifs of networks induced
  from time series}},
\newblock \bibinfo{journal}{P. Natl. Acad. Sci. USA} \bibinfo{volume}{105}
  (\bibinfo{year}{2008}) \bibinfo{pages}{19601}.
%Type = Book
\bibitem[{Collet and Eckmann(2006)}]{Collet06}
\bibinfo{author}{P.~Collet}, \bibinfo{author}{J.~Eckmann},
  \bibinfo{title}{{Concepts and results in chaotic dynamics: a short course}},
  \bibinfo{publisher}{Springer Verlag}, \bibinfo{year}{2006}.
%Type = Inproceedings
\bibitem[{Dellnitz et~al.(2006)Dellnitz, Molo, Metzner, Preis, and
  Sch{\"u}tte}]{Dellnitz06}
\bibinfo{author}{M.~Dellnitz}, \bibinfo{author}{M.~Molo},
  \bibinfo{author}{P.~Metzner}, \bibinfo{author}{R.~Preis},
  \bibinfo{author}{C.~Sch{\"u}tte},
\newblock \bibinfo{title}{{Graph algorithms for dynamical systems}},
\newblock in: \bibinfo{booktitle}{Analysis, modeling and simulation of
  multiscale problems}, \bibinfo{publisher}{Springer}, \bibinfo{year}{2006},
  pp. \bibinfo{pages}{619--645}.
%Type = Article
\bibitem[{Dellnitz and Junge(1999)}]{Dellnitz99}
\bibinfo{author}{M.~Dellnitz}, \bibinfo{author}{O.~Junge},
\newblock \bibinfo{title}{On the approximation of complicated dynamical
  behavior},
\newblock \bibinfo{journal}{SIAM Journal on Numerical Analysis}
  \bibinfo{volume}{36} (\bibinfo{year}{1999}) \bibinfo{pages}{491--515}.
%Type = Article
\bibitem[{Froyland and Dellnitz(2003)}]{Froyland03}
\bibinfo{author}{G.~Froyland}, \bibinfo{author}{M.~Dellnitz},
\newblock \bibinfo{title}{{Detecting and locating near-optimal almost-invariant
  sets and cycles}},
\newblock \bibinfo{journal}{SIAM J. Sci. Comput.} \bibinfo{volume}{24}
  (\bibinfo{year}{2003}) \bibinfo{pages}{1839--1863}.
%Type = Book
\bibitem[{Chapelle et~al.(2006)Chapelle, Sch{\"o}lkopf, and Zien}]{Chapelle06}
\bibinfo{editor}{O.~Chapelle}, \bibinfo{editor}{B.~Sch{\"o}lkopf},
  \bibinfo{editor}{A.~Zien} (Eds.), \bibinfo{title}{Semi-Supervised Learning},
  \bibinfo{publisher}{MIT Press}, \bibinfo{address}{Cambridge, MA},
  \bibinfo{year}{2006}.
%Type = Book
\bibitem[{Hastie et~al.(2009)Hastie, Tibshinari, and Freedman}]{Hastie09}
\bibinfo{author}{T.~Hastie}, \bibinfo{author}{R.~Tibshinari},
  \bibinfo{author}{J.~Freedman}, \bibinfo{title}{The elements of statistical
  learning}, \bibinfo{publisher}{Springer Verlag}, \bibinfo{year}{2009}.
%Type = Article
\bibitem[{Freedman(1966)}]{freedman66}
\bibinfo{author}{H.~Freedman},
\newblock \bibinfo{title}{{``The Little Variable Factor" A Statistical
  Discussion of the Reading of Seismograms}},
\newblock \bibinfo{journal}{Bull. seism. Soc. Am.} \bibinfo{volume}{56(2)}
  (\bibinfo{year}{1966}) \bibinfo{pages}{593--604}.
%Type = Techreport
\bibitem[{Velasco et~al.(2001)Velasco, Young, and Anderson}]{Velasco01}
\bibinfo{author}{A.~Velasco}, \bibinfo{author}{C.~Young},
  \bibinfo{author}{D.~Anderson}, \bibinfo{title}{Uncertainty in Phase Arrival
  Time Picks for Regional Seismic Events: An Experimental Design},
  \bibinfo{type}{Technical Report}, US Department of Energy,
  \bibinfo{year}{2001}.
%Type = Article
\bibitem[{Anant and Dowla(1997)}]{Anant97}
\bibinfo{author}{K.~Anant}, \bibinfo{author}{F.~Dowla},
\newblock \bibinfo{title}{{Wavelet transform methods for phase identification
  in three-component seismograms}},
\newblock \bibinfo{journal}{Bull. seism. Soc. Am.} \bibinfo{volume}{87}
  (\bibinfo{year}{1997}) \bibinfo{pages}{1598}.
%Type = Article
\bibitem[{Gendron et~al.(2000)Gendron, Ebel, and Manolakis}]{Gendron00}
\bibinfo{author}{P.~Gendron}, \bibinfo{author}{J.~Ebel},
  \bibinfo{author}{D.~Manolakis},
\newblock \bibinfo{title}{{Rapid joint detection and classification with
  wavelet bases via Bayes theorem}},
\newblock \bibinfo{journal}{Bull. seism. Soc. Am.} \bibinfo{volume}{90}
  (\bibinfo{year}{2000}) \bibinfo{pages}{764}.
%Type = Article
\bibitem[{Schclar et~al.(2010)Schclar, Averbuch, Rabin, Zheludev, and
  Hochman}]{Schclar10}
\bibinfo{author}{A.~Schclar}, \bibinfo{author}{A.~Averbuch},
  \bibinfo{author}{N.~Rabin}, \bibinfo{author}{V.~Zheludev},
  \bibinfo{author}{K.~Hochman},
\newblock \bibinfo{title}{{A diffusion framework for detection of moving
  vehicles}},
\newblock \bibinfo{journal}{Digit. Signal Process.} \bibinfo{volume}{20}
  (\bibinfo{year}{2010}) \bibinfo{pages}{111--122}.
%Type = Article
\bibitem[{Arya et~al.(1998)Arya, Mount, Netanyahu, Silverman, and Wu}]{Arya98}
\bibinfo{author}{S.~Arya}, \bibinfo{author}{D.~Mount},
  \bibinfo{author}{N.~Netanyahu}, \bibinfo{author}{R.~Silverman},
  \bibinfo{author}{A.~Wu},
\newblock \bibinfo{title}{{An optimal algorithm for approximate nearest
  neighbor searching fixed dimensions}},
\newblock \bibinfo{journal}{J. ACM} \bibinfo{volume}{45} (\bibinfo{year}{1998})
  \bibinfo{pages}{891--923}. \bibinfo{note}{(available online at
  http://www.cs.umd.edu/\Pisymbol{psy}{126}mount/ANN/)}.
%Type = Article
\bibitem[{Deuflhard et~al.(2000)Deuflhard, Huisinga, Fischer, and
  Sch{\"u}tte}]{Deuflhard00}
\bibinfo{author}{P.~Deuflhard}, \bibinfo{author}{W.~Huisinga},
  \bibinfo{author}{A.~Fischer}, \bibinfo{author}{C.~Sch{\"u}tte},
\newblock \bibinfo{title}{{Identification of almost invariant aggregates in
  reversible nearly uncoupled Markov chains}},
\newblock \bibinfo{journal}{Linear Algebra Appl.} \bibinfo{volume}{315}
  (\bibinfo{year}{2000}) \bibinfo{pages}{39--59}.
%Type = Article
\bibitem[{Horenko(2008)}]{Horenko08}
\bibinfo{author}{I.~Horenko},
\newblock \bibinfo{title}{{On simultaneous data-based dimension reduction and
  hidden phase identification}},
\newblock \bibinfo{journal}{J. Atmos. Sci.} \bibinfo{volume}{65}
  (\bibinfo{year}{2008}) \bibinfo{pages}{1941--1954}.
%Type = Article
\bibitem[{Broomhead et~al.(1996)Broomhead, Huke, and Potts}]{Broomhead96}
\bibinfo{author}{D.~Broomhead}, \bibinfo{author}{J.~Huke},
  \bibinfo{author}{M.~Potts},
\newblock \bibinfo{title}{{Cancelling deterministic noise by constructing
  nonlinear inverses to linear filters}},
\newblock \bibinfo{journal}{Physica D} \bibinfo{volume}{89}
  (\bibinfo{year}{1996}) \bibinfo{pages}{439--458}.
%Type = Article
\bibitem[{Kostelich and Schreiber(1993)}]{Kostelich93}
\bibinfo{author}{E.~Kostelich}, \bibinfo{author}{T.~Schreiber},
\newblock \bibinfo{title}{{Noise reduction in chaotic time-series data: a
  survey of common methods}},
\newblock \bibinfo{journal}{Phys. Rev. E} \bibinfo{volume}{48}
  (\bibinfo{year}{1993}) \bibinfo{pages}{1752--1763}.
%Type = Article
\bibitem[{Casdagli(1989)}]{Casdagli89}
\bibinfo{author}{M.~Casdagli},
\newblock \bibinfo{title}{{Nonlinear prediction of chaotic time series}},
\newblock \bibinfo{journal}{Physica D} \bibinfo{volume}{35}
  (\bibinfo{year}{1989}) \bibinfo{pages}{335--356}.
%Type = Book
\bibitem[{Kantz and Schreiber(2004)}]{Kantz04}
\bibinfo{author}{H.~Kantz}, \bibinfo{author}{T.~Schreiber},
  \bibinfo{title}{{Nonlinear time series analysis}},
  \bibinfo{publisher}{Cambridge University Press}, \bibinfo{year}{2004}.

\end{thebibliography}

\end{document}